\documentclass{appolb}
\usepackage{graphicx}
\usepackage{amsmath}

\def\vec#1{\ensuremath{\mathchoice
                     {\mbox{\boldmath$\displaystyle#1$}}
                     {\mbox{\boldmath$\textstyle#1$}}
                     {\mbox{\boldmath$\scriptstyle#1$}}
                     {\mbox{\boldmath$\scriptscriptstyle#1$}}}}

\begin{document}
%
\title{Dirac shell quark-core model 
for the study of  non-strange baryonic spectroscopy
}
\author{M. De Sanctis \footnote{mdesanctis@unal.edu.co}
\address{Universidad Nacional de Colombia, Bogot\'a, Colombia }
\\
}
\maketitle
\begin{abstract}
A Dirac shell model is developed for the study of baryon spectroscopy, 
taking into account the most relevant results of the quark-diquark models.
%
The lack of translational invariance
of the shell model
is avoided, in the present work, by introducing
a scalar-isoscalar fictitious particle that represents the origin of quark shell interaction;
in this way the states of the system are eigenstates  of the total momentum of the baryon.
Only one-particle excitations are considered.
A two-quark core takes the place of the diquark, while the  third quark is excited to reproduce the baryonic resonances.
For the $N(939)$ and $\Delta(1232)$, that represent the ground states of the spectra, the three quarks are considered
identical particles and the wave functions are completely antisymmetric. 
%
The model is used to calculate the  spectra of the  $N$ and $\Delta$ resonances and the nucleon magnetic moments. 
The results are compared to the present experimental data. 
Due to the presence of the core and to the one-particle excitations, the structure of the  obtained spectra is analogous
to that given by the quark-diquark models.
\end{abstract}
\PACS{
      {12.39.Ki},~~
      {12.39.Pn},~~
      {14.20.Gk}
     } 
\section{Introduction}
Due to the great difficulties to solve directly the field equations of Quantum Chromo-Dynamics (QCD),
light baryon spectroscopy has been widely studied by means of a variety of different quark models.
Without any attempt to be exhaustive but only with the aim of introducing some relevant concepts for the development of the present work,
we tentatively group these models in the following way:
single particle relativistic models (SPRMs), constituent quark models (CQMs) and quark-diquark models (QDMs).

In general, in the SPRMs,  the independent motion of the three quarks is considered, so that the 
wave function of the system is given  by the product of three single particle wave functions,
where each  wave fuction represents the state of a quark.
In particular, we recall the historically relevant MIT relativistic bag models (RBMs)
and the relativistic chiral (shell) model (RCM).
In the RBMs
\cite{bag1,bag2,bag3,bag4,bag5,bag6,bag7,bag8}
three massless (or light) quarks are moving inside a spherical bag where a field energy density is also present.
In consequence, the quark wave functions are given by standard  Dirac spinors with spherical Bessel functions.
In the original formulation \cite{bag1,bag2,bag3,bag4},
energy quantization is obtained by means of a boundary condition 
that takes into account the energy-momentum conservation at the surface of the bag. 
A residual quark-quark vector interaction is introduced to remove the degeneracy between the $N(939)$ and the $\Delta(1232)$.
A good description of the ground state properties was achieved but the reproduction of the excited state spectra was less accurate.
From a fundamental point of view, the RBMs, as all the SPRMs,  are not translationally invariant, so that the total wave function
does not represent an eigenstate of the total momentum of the baryon. 
\textit{ Ad hoc} procedures are used to subtract from the total energy the spurious contributions of the center of mass motion
\cite{bag5,bag9,bag10}.

A RCM was proposed  in which   the pionic field is explicitly introduced 
\cite{chir1}.
Moreover, the valence quark interaction  is based on the one-pion exchange mechanism. 
That model contains some ideas that, as it will be explained in the following, have been also used to develop the  present work.
In particular, in ref. \cite{chir1}, the author makes the hypothesis that two quarks belong to the ground S-wave shell 
while the third quark goes to the excited shells in order to reproduce the baryonic spectra.
However, the experimental energies of the resonances are not reproduced with high accuracy because no extra quark-quark  interaction is introduced.
As in the case of  RBMs, a particular technique is used to subtract the spurious effects related to the center of mass motion.  
Subsequently, the same author also developed a field theoretical model in which the translational invariance is completely satisfied 
from the beginning
\cite{chir2}.

The CQMs have represented a very successfull method for the study of light baryon spectrum  
\cite{isg:1,isg:2,isg:3,Capstick:1986bm,bij:1,bij:2,blr:1n1,sig,bis,blr:1n2,blr:1n3,blr:1n4,blr:2n1,blr:2n2,blr:2n3,blr:2n4,lor:1,lor:2}.
In these theoretical models
baryons are described as  bound states of three constituent quarks 
that can be considered as effective degrees of freedom  representing
the three valence quarks inside baryons, \textit{dressed} by  virtual gluons and $q \bar q$ sea pairs.
In consequence, for their mass, a much higher value than the QCD \textit{current mass} is generally taken.
The spatial dynamical variables that are used to study the three quark systems 
in the center of mass reference frame, are the Jacobi coordinates $\vec \rho$ and $\vec \lambda$,
where the former represents the distance between the first two quarks and the latter  the distance
between the third quark and the center of mass of the first two.
In this way the CQMs are translationally invariant and the wave function is an eigenstate of the total momentum of the baryon.
We recall that the  total wave function is written as  the product of two factors:
the \textit{first factor} is given by a sum of products  of spatial, spin and isospin terms; 
the \textit{second factor} represents the antisymmetric (white) color  term. 
In consequence the 
\textit{first factor} must be symmetric with respect to the exchange of every pair of quarks. 
Furthermore, the use of the Jacobi variables gives rise to a quite complex (in any case, not ``independent particle'')  
structure for the spatial terms of the quark-quark interaction.
Due to this complexity, CQMs have been formulated initially by means of nonrelativistic or relativized Hamiltonians.
A fully relativistic study by means of Dirac spinors was given in the model of refs. \cite{lor:1,lor:2}.
However, also in this model, the quark mass is of the order of $300~ MeV$, which value is, in any case, much higher than the QCD current quark mass.

\noindent
In general, in the CQMs, the light baryons can  be ordered according to the approximate SU$_{\mbox{f}}$(3) symmetry into the multiplets 
$[{\bf 1}]_A \oplus [{\bf 8}]_M \oplus [{\bf 8}]_M \oplus [{\bf 10}]_S$. 
CQMs reproduce with good accuracy several properties of baryons, such as the strong decays, the magnetic moments
and the electromagnetic elastic form factors.
However, they predict a larger number of states than the experimentally observed resonances, 
that is known as \textit{the missing resonance problem}. 
Furthermore, some states with certain quantum numbers appear in the spectrum at excitation 
energies much lower than predicted \cite{Nakamura:2010zzi}. 
The problem of the missing resonances \cite{Nakamura:2010zzi,Capstick:1992uc,Capstick:1992th} has motivated the realization of 
several experiments, such as CB-ELSA 
\cite{blr:3n1}, CBELSA/TAPS \cite{blr:3n2}, TAPS \cite{blr:4n1,blr:4n2,blr:4n3}, GRAAL \cite{graal:1,graal:2}, 
SAPHIR \cite{saphir:1,saphir:2} 
and CLAS \cite{clas:1,clas:2,clas:3}, which only provided a few weak indications about some states. 
Even though several experiments have been dedicated to the search of missing resonances, just a small number of 
them has been included into the resonance list \cite{Nakamura:2010zzi}.

Three possible solutions have been proposed for the missing resonance problem: 

\noindent
1) considering the detection mechanism, some resonances may be very wea\-kly 
coupled to the single pion, but with higher probabilities of decaying into 
two or more pions or into other mesons \cite{Nakamura:2010zzi,Capstick:1992uc,Capstick:1992th};
further difficulties can be given by the problem of the separation of the experimental data from 
the background and by the expansion of the differential cross sections into many partial waves;

\noindent
2) theoretically, it is possible to construct effective models that are characterized by a smaller number of \textit{active}  degrees of freedom 
with respect to   the three quarks of the CQMs;
in this way the majority of the missing resonances, not yet experimentally observed, are simply \textit{not predicted} by these models;

\noindent
3) only a selected set of excited states are retained; in particular, the \textit{one-particle} excited states, are taken to represent the experimental
spectra of light baryons; this choice, proposed in ref. \cite{chir1},  is  made also in the present work.

\noindent
We highlight that  the solution (2) represents the basic assumption of the widely developed QDMs
\cite{blr:5n1,blr:5n2,Anselmino:1992vg,Jaffe:2004ph,Wilczek:2004im,Selem:2006nd,Santopinto:2004hw,Forkel:2008un,anis:1,anis:2,ferr:1,ferr:2,DeSanctis:2011zz,Galata:2012xt,col_last}
that are able to reproduce the baryonic spectra with  high accuracy.

\noindent
The notion of diquark dates back to 1964, when its possibility was mentioned by Gell-Mann \cite{GellMann:1964nj} 
in his original paper on quarks. 
Since then, many papers have been written on this topic (for a review see 
Ref. \cite{Anselmino:1992vg}) and, more recently, the diquark concept has been applied to various calculations 
\cite{Jaffe:2004ph,Wilczek:2004im,Selem:2006nd,Santopinto:2004hw,Forkel:2008un,anis:1,anis:2,ferr:1,ferr:2,DeSanctis:2011zz,Galata:2012xt,Jakob:1997,Bloch:1999ke,Brodsky:2002,Ma,Oettel:2002wf,Gamberg:2003,Jaffe:2003sg,Maris:2004,DeGrand:2007vu,BacchettaRadici}.

\noindent
For  the present study we only recall that,
in ref. \cite{Santopinto:2004hw}, it was developed an nonrelativistic interacting quark-diquark mo\-del, 
\textit{i.e.} a potential model based on the effective degrees of freedom of a constituent quark and diquark. 
In refs. \cite{ferr:1,ferr:2}, it was  ``relativized''   and reformulated within the Point Form formalism \cite{pf1,pf2,pf3}.
In ref. \cite{DeSanctis:2011zz}, the wave functions of refs. \cite{ferr:1,ferr:2} were used to compute the nucleon electromagnetic 
form factors.
An accurate reproduction of the baryonic spectra was obtained in
a relativistic QDM in which a spin-isospin transition interaction was introduced with the aim of mixing the scalar and the axial-vector diquarks \cite{col_last}.
We shall consider mainly
that work for a comparison with the results of the present model.

\noindent
We point out that in the QDMs
the effective degree of freedom of the diquark is introduced to describe baryons as bound states 
of a constituent diquark and a quark \cite{blr:5n1,blr:5n2}.
In more detail, two quarks are supposed to be strongly correlated (say, \textit{frozen})  in the constituent diquark;
their relative motion is assumed to have a vanishing relative orbital angular momentum, that is $L_d=0$.
The diquark can be found in two orthogonal states of spin $S_d$ and isospin $T_d$: the \textit{scalar diquark} with $S_d=T_d=0$ and the \textit{axial-vector diquark} with 
$S_d=T_d=1$.
These quantum numbers are determined considering that: 
\begin{itemize}
\item
the \textit{frozen} quarks of the diquark are identical particles that satisfy the Pauli exclusion
principle; 
\item
the color factor is given by the standard antisymmetric (white) function, as in CQMs.
\end{itemize}

\noindent
On the other hand, the motion of the quark with respect to the diquark is described by the spatial variable $\vec r$ that represents their relative
distance.
The use of only one spatial variable ($\vec r$) instead of the two variables ($\vec \rho$ and $\vec \lambda$) of the CQMs
gives rise to a number of excited states that is \textit{substantially reduced} with respect to the predictions of the three quark CQMs.
Furthermore, the obtained spectrum has a \textit{one-particle} excitation structure, more consistent with the experimental data.

In all the previously mentioned investigations, the  diquarks were used as effective degrees of freedom of the baryonic states.
However, within the QDMs, the dynamical mechanism that allows for the diquark formation is not specified:
the diquark is directly assumed as a new effective particle with the same quantum numbers of two strongly correlated quarks.
In consequence, the wave function is not antisymmetrized with respect to the interchange of a quark \textit{belonging} to the diquark
and of the \textit{external} quark.
\vskip 0.5 truecm

These theoretical difficulties motivated the development of the present model
with the objective of reproducing the same structure of the QDM spectra, that is  a one-particle excitation structure
and no missing resonances.
However, in our model the three quarks are present as \textit{real} degrees of freedom.
In more detail, we construct a quasi-independent particle shell model in which the resonances of the spectra
are given by the excitation of one quark while the other two quarks always remain in the first  shell 
with vanishing orbital angular momentum.
These two quarks form a \textit{core}, that replaces the diquark of the QDMs. 
In particular, due to the antisymmetry of the wave function, the two quarks of the core
(analogously to the diquark case)
can be found in two orthogonal states: the state with $S_c=T_c=1$ and the state with $S_c=T_c=0$,
where $S_c$ and $T_c$ respectively represent  spin and isospin of the core.
Above we have used the definition of ``quasi-independent'' particle model because, as it will be explained in the following,
we introduce a fictitious particle in order to obtain a translational invariant model. 
In consequence,
the kinetic energy  associated to that particle  does  give rise to a  non-independent particle operator that,
however,
will be treated perturbatively.

\vskip 0.5 truecm

In the following sect. \ref{general} we shall give an overall description of the model.
In sect. \ref{model}, starting with the definition of the spatial variables, 
we shall formally construct 
the Hamiltonian of the model.
In sect. \ref{mag_mom} we shall explain the calculation of the  magnetic moments of the proton and neutron.
In sect. \ref{results} the results for the spectra and magnetic moments will be shown and commented.
Finally, in sect. \ref{conclusion}, some conclusions will be drawn and some possible perspectives will be illustrated.

The technical details of the model will be analyzed in the appendices.
In app. \ref{app_dir}, the main properties of the one-body Dirac equation with spin symmetry will be discussed.
In app. \ref{app_ho}, the same formalism will be applied to the case of a harmonic oscillator interaction.
The magnetic dipole operator, for the spin symmetry one-body Dirac equation, will be studied in app. \ref{app_magdip}.
Finally, in app. \ref{app_wf}, the three quark complete wave functions will be constructed and the numerical procedure for the solution will
be synthetically described. For the calculations we use the so-called natural units, that is $\hbar=c=1$.

\vskip 1.0 truecm

\section{General description of the model}\label{general}
In this section we  discuss, at a general level,  the different parts of the model with respect to its objectives, 
also making a critical comparison with the choices of other studies;
on the other hand, more details about the formulation of the model  will be explained in the next section \ref{model}. 

\vskip 0.5 truecm
\noindent
The construcion of the model is based on the five points that are illustrated in the following.

\noindent
i) The effective particles of the model are represented by three  very \textit{light quarks} and a 
fictitious scalar-isoscalar particle, denoted, in the following,  as  ``$x$-particle''.
Pictorially, we can say that in our model the baryon 
``looks like''  a Lithium atom, in which  the electrons are replaced by the quarks
and the nucleus  is replaced by and the $x$-particle. 
(To avoid misunderstanding, note that, in the ground states, the three quarks belong to the same shell,
while this configuration  is forbidden for the  electrons of the Lithium atom).


\noindent
In a pure shell model the quark interaction would be referred to the origin of the coordinates, violating the translational invariance.
This difficulty is avoided here assuming that the quark interaction depends on the distance between the quark and
the $x$-particle.
Obviously, the $x$-particle possesses a momentum and a kinetic energy.
In the present model,
its momentum, in the Center of Mass (CM) reference frame of the baryon, is \textit{opposite}
to the \textit{sum} of the quark momenta; in consequence, the baryon state is an eigenstate of the total momentum;
in particular, the total momentum is zero in the CM.
For the kinetic energy of the $x$-particle, in the case of a sufficiently high mass, a nonrelativistic  expansion  can be performed
and the contributions of this term can be calculated \textit{perturbatively}, without spoiling the independent particle character of the model.
For this reason, we introduced above  the definition of ``quasi-independent'' particle model.
Furthermore, the mass of the $x$-particle also represents, in this work, the zero point energy of the spectrum.
The contribution of the kinetic energy of the $x$-particle will be studied in subsect. \ref{k_h_x}.

\noindent
Without attempting to attribute a \textit{real} character to the $x$-particle, we recall that
the hypothesis of an effective bosonic particle (the so-called pomeron) is not unusual
in the study of other problems of hadronic physics. 
It was introduced to study baryonic scattering, also  
in the framework of QCD; see, for example, ref. \cite{pom1}.
Some works have identified the pomeron as a tensorial particle \cite{pom2,pom3}
and  a model has been proposed in which it is represented as bound state of two effective gluons \cite{pom4}.\\
To avoid confusion, we give here a brief  terminological explanation: we shall use the term \textit{core} (introduced above for the two quarks of the first shell)
without including the $x$-particle, 
for the two following reasons: 
\textit{i}) in  the present work, the $x$-particle is essentially considered as a \textit{fictitious} particle;
\textit{ii}) in any case, it interacts also with the quark not belonging to the \textit{core}.
We also note that the $x$-particle does not bring angular momentum that, as it will be explained in the following,
is brought by the two quarks of the core and by the third quark.
  
\vskip 0.5 truecm
\noindent
ii) Another relevant objective of our model is the use of very \textit{light quarks}
without appealing for a mechanism that generates the constituent mass.
More precisely, we consider
the standard value of the ``current-quark mass'' that is estimated by means of  a mass-independent subtraction scheme
in the QCD theory.
In particular, for this model we take the mean value of the up and down quark, that is $m_q=(m_u+m_d)/2=3.5 ~ MeV$ \cite{pdg}.

\noindent
Due to this hypothesis, the quark motion is  extremely relativistic. 
Correspondingly, the formulation of a model with a  three-body ultrarelativistic equation, suitable for this choice, would involve a high level of complexity.
This is another argument (beyond the structure of the spectrum) to prefer, at this stage, a quasi-single-particle model
where the quark motion is  described, in a first approximation, by independent Dirac equations.

\vskip 0.5 truecm
\noindent
iii) As for the interaction in the Dirac equation, we take two  central terms of \textit{equal magnitude}:
a scalar term and the zero component of a vectorial interaction.
With this choice, that corresponds to the so-called ``spin symmetry case'' \cite{spsya,spsyb,spsyc,spsyd,spsye,spsyf,spsyg},
the quark orbital angular momentum and its spin are decoupled, so that no spin-orbit interaction is produced.
Due to this property, Dirac equation with spin symmetry was used to study mesonic spectra \cite{spsyc,spsye,spsyg}.
Also in the present case of baryonic spectra, this option is strongly favored by the experimental data
that show only a very small spin-orbit splitting of the baryonic resonances.
Moreover, we point out  that, theoretically,  in QDMs, the spin-orbit
interaction is usually neglected in a first approximation; see, for example, ref. \cite{col_last}.

\noindent
Another interesting property of the Dirac equation with spin symmetry is that
it can be  transformed into
a Schr\"o\-din\-ger-like,  energy-dependent equation, reducing the numerical complexity of the solution procedure.
In the pre\-sent work we take a harmonic oscillator interaction to represent the main contribution to quark confinement; 
we obtain, in this way, an analytically solvable equation.
Other contributions, all with spin symmetry, are added to reproduce in more  detail the structure of the spectra.
The Dirac quark Hamiltonian is introduced in subsect. \ref{dirac_term}. 
The Dirac equation with spin symmetry is studied in app. \ref{app_dir} 
and specialized to the harmonic oscillator interaction in app. \ref{app_ho}.

\noindent
The present choice of a Dirac equation with spin symmetry strongly differs from that of ref. \cite{chir1}
where a pseudoscalar interaction  related to one-pion exchange was considered.

\noindent
A phenomenological spin-spin interaction is also introduced to remove the degeneracy between the $N(939)$ and the $\Delta(1232)$
and to reproduce in more detail the resonance levels.
For simplicity, also in this case, the spatial dependence is taken as a central function of the quark distance with respect 
to the position of the $x$-particle; for this interaction, see subsect. \ref{hss}.

\vskip 0.5 truecm
\noindent
iv) We now discuss  the implementation in the model of the Pauli exclusion principle that implies the 
antisymmetric character of the  quark wave function.
As seen before, 
this principle  is considered as a basic assumption
in the SPRMs and  CQMs. 
On the other hand,
in QDMs, the quarks inside the diquark do not appear as dynamical degrees of freedom,
so that, for the quark outside the diquark,  no antisymmetrization is required.

\noindent
%
For the ground states, \textit{i. e.} the  $N(939)$ and the $\Delta(1232)$,
within the present model
(in which all the tree quarks belong to the first shell)
no reason can be found (within the model) to refuse this basic principle.
In this sense, we recall that,
historically, it compelled the introduction of the color quantum number, when applied to the
wave function of the $\Delta(1232)$.

\noindent
Let us analyze 
the case of  $N(939)$.
Assuming, in a standard way, that the spatial term of the wave function is symmetric,
the total antisymmetry of the wave function requires a symmetric  spin-isospin factor.
This symmetric factor must contain the core states 
%
$|S_c=T_c=0>$ and  $|S_c=T_c=1>$ with \textit{equal amplitudes},
that is $a_0=a_1=1/\sqrt{2}$.
(We have used $S_c$ and $T_c$ to denote, as before, the spin and the isospin of the pair of quarks $1$ and $2$ belonging to the core).

\noindent
Generally, in QDMs (where the total antisymetrization is not required) the two amplitudes can be not equal;
in some QDMs, see for example ref. \cite{ferr:1}, the amplitude of the state $|S_c=T_c=1>$  is vanishing.
But, if a spin-isospin transition interaction is introduced into the dynamics of the model, we highlight that
the solution of the eigenvalue equation gives two amplitudes having very similar values, that is
$a_0=(a_S)=0.727~,~~~a_1=(a_V)=0.687$  \cite{col_last}, suggesting that, also in QDMs, a symmetric spin-isospin factor
can be a good approximation for the $N(939)$ wave function.

\noindent
The case of the $\Delta(1232)$ is even more obvious: the spatial, spin and isospin factors must be, \textit{all}, symmetric
with respect to quark interchange.

\noindent
Concluding, in the present work,  we consider the three quarks, in the ground states,   as identical
particles, with the standard consequences, discussed above, for the wave functions.
These wave functions will be given explicitly in  eqs. (\ref{Phi_N_a}), (\ref{Phi_N_b}) and (\ref{psi_gen_fact}), for $N(939)$ and $\Delta(1232)$, respectively.

\noindent
On the other hand, considering that in our model only \textit{one-quark} excitations are taken into account,
we make the same hypothesis of the QDMs:  the excited quark is considered \textit{not necessarily}
identical to the two quarks of the core.
This assumption gives the correct spectroscopy, analogously to  the QDMs, with no missing resonances.
Phenomenologically, this assumption
can be justified observing that the excited quark is in a different energy state with respect 
to the two quarks of the core.
In consequence, its \textit{effective} properties, in particular the interaction, 
are modified by the strong field and its effective interaction 
is different  with respect to the interaction of the two quarks
of the core.
For this reason, we shall take different parametrizations for the interaction of the quarks in the core and 
for the interaction of the excited quark.

\vskip 0.5 truecm
\noindent
v) With the assumptions discussed above, we can introduce here the basic structure of the spectroscopy of the model.
Preliminarly, we assign the indices $1$ and $2$ to the two quarks of the core and the index $3$ to the quark that can be excited
to higher levels.

\noindent
In the first place, we analyze the coupling scheme for the angular momenta.
For the orbital angular momentum of the core, we have:
\begin{equation}\label{l_core}
\vec L_c=\vec l_1+ \vec l_2~.
\end{equation}
For the total orbital angular momentum, one has:
\begin{subequations}
\begin{equation}\label{l_a_tot}
 \vec L= \vec L_c +   \vec l_3~.
\end{equation}
However, given that we always have $L_c=0$,
the total orbital angular momentum simply is:
\begin{equation}\label{l_tot}
\vec L= \vec l_3~.
\end{equation}
\end{subequations}
%

\noindent
The core spin is:
\begin{equation}\label{s_core}
\vec S_c=\vec s_1+ \vec s_2~
\end{equation}
being $S_c=0,1$;
the total spin is:
\begin{equation}\label{s_tot}
\vec S= \vec S_c+ \vec s_3~;
\end{equation}
the possible values for $S$ are $S=1/2,~3/2$.

\noindent
Finally, the total angular momentum is:
\begin{equation}\label{j_tot}
\vec J= \vec L + \vec S~.
\end{equation}
For the isospin of the core we have:
\begin{equation}\label{t_core}
\vec T_c=\vec t_1+ \vec t_2~,
\end{equation}
being $T_c=0,1$.
As discussed above, the Pauli exclusion priciple requires $S_c=T_c$;
the total isospin is:
\begin{equation}\label{t_tot}
\vec T= \vec T_c+ \vec t_3
\end{equation}
and the possible values for $T$ are $T=1/2$ ($N$ states), and $T=3/2$ ($\Delta$ states);
these latter states only have $S_c=T_c=1$.

\noindent
We also introduce the parity of the state:
\begin{equation}\label{parity}
P=(-1)^L
\end{equation}
and, finally,  $n_r=0,1,2,....$ that represents the  radial  excitation number  of the quark $3$.

\noindent
The states are identified by  the following notation:
\begin{equation}\label{psi}
 |\Psi>=| T; n_r, L, S_c, S, J^P  >  
\end{equation}
where, for simplicity, the ``third components'' $M_T$ and $M_J$ have been omitted.
\vskip 0.5 truecm

\noindent
We now consider the list of the ``first'' 
states  of the model;
their quantum numbers are displayed in table (\ref{tab:nfirststates}) and  table (\ref{tab:deltafirststates}), for the $N$
and $\Delta$ spectrum, respectively.
We have taken the states with $L\leq2$. For $L=0$,  we have taken  $n_r=0,~1$; for $L=1,2$, $n_r=0$ only. 
With the previous choices, we have  taken all the possible values for $S_c, S$ and $J$.
For the case on the $N$ resonances
we  have also considered one state with $n_r=2$ (the last of table (\ref{tab:nfirststates})) in order  to reproduce the  $N(1880) {\frac 1 2}^+$.
\noindent
%
\noindent
The excitation energies of the states of table  (\ref{tab:nfirststates}) and table (\ref{tab:deltafirststates})  
roughly correspond to the energies of the states with $N \leq 2$ in the standard  CQMs \cite{pdg}
but our model predicts \textit{less} states than the CQMs.

\begin{table}[h]  
\caption{ Quantum numbers of the first $N$ states of the model.}
\begin{center}
\begin{tabular}{lllll}
\hline
 $n_r$  &$L$   &$S_c$   &$S$        &$J^P$  \\ 
\hline \\
$0$     & $0$  &$(0,1)$   & $1/2$     & $1/2^+$ \\
$1$     & $0$  &$0$     & $1/2$     & $1/2^+$ \\
$1$     & $0$  &$1$     & $1/2$     & $1/2^+$ \\
$1$     & $0$  &$1$     & $3/2$     & $3/2^+$ \\
$0$     & $1$  &$0$     & $1/2$     & $1/2^- ~ 3/2^-$ \\
$0$     & $1$  &$1$     & $1/2$     & $1/2^- ~ 3/2^-$ \\
$0$     & $1$  &$1$     & $3/2$     & $1/2^- ~ 3/2^- ~ 5/2^-$ \\
$0$     & $2$  &$0$     & $1/2$     & $3/2^+ ~ 5/2^+$\\
$0$     & $2$  &$1$     & $1/2$     & $3/2^+ ~ 5/2^+$\\
$0$     & $2$  &$1$     & $3/2$     & $1/2^+ ~ 3/2^+ ~ 5/2^+ ~7/2^+$\\
$2$     & $0$  &$0$     & $1/2$     & $1/2^+$ \\
\hline
\hline
\end{tabular}
\end{center}
\label{tab:nfirststates}
\end{table}

\begin{table}[h]  
\caption{ Quantum numbers of the first $\Delta$ states of the model.}
\begin{center}
\begin{tabular}{lllll}
\hline
 $n_r$  &$L$   &$S_c$   &$S$        &$J^P$  \\ 
\hline \\
$0$     & $0$  &$1$     & $3/2$     & $3/2^+$ \\
$1$     & $0$  &$1$     & $1/2$     & $1/2^+$ \\
$1$     & $0$  &$1$     & $3/2$     & $3/2^+$ \\
$0$     & $1$  &$1$     & $1/2$     & $1/2^-~ 3/2^-$ \\
$0$     & $1$  &$1$     & $3/2$     & $1/2^- ~ 3/2^- ~5/2^- $ \\
$0$     & $2$  &$1$     & $1/2$     & $3/2^+ ~ 5/2^+$ \\
$0$     & $2$  &$1$     & $3/2$     & $1/2^+ ~ 3/2^+ ~ 5/2^+ ~7/2^+$ \\

\hline
\hline
\end{tabular}
\end{center}
\label{tab:deltafirststates}
\end{table}

\noindent
The previous states will be used to reproduce 
the experimental baryonic spectra, without missing resonances, up to $2000~MeV$.

\noindent
The mass values of each state will be determined by the model calculations.
Being absent a spin-orbit interaction, the states with different $J$ but with the same values for the other
quantum numbers are degenerate.
The first state of table \ref{tab:nfirststates} and that of table \ref{tab:deltafirststates}
respectively represent the
$N(939)$ and the $\Delta(1232)$.
As discussed above, the $N(939)$ wave function is completely antisymmetric,
requiring $S_c=0$ and $S_c=1$ (with equal amplitudes) as it is indicated in the first line of table \ref{tab:nfirststates}.

\vskip 2.5 truecm

\section{Construction of the Hamiltonian of the model}\label{model} 

\subsection{The coordinates and conjugate momenta}\label{coord_mom}
The first task is to define  the coordinates and the conjugate momenta of the constituents of the model.
In a generic frame, we introduce  the coordinates $\vec x_i$, $\vec x_x$ that respectively represent the position 
of the three quarks ($i=1,2,3$) and of the $x$-particle. 
The corresponding canonical conjugate momenta are $\vec k_i$, $\vec k_x$.
The three quarks have equal mass $m_q$; the $x$-particle mass is $m_x$.

\noindent
We now define the intrinsic coordinates $\vec r_i$, that will be used in the calculation,
and the position of the center of mass $\vec R$, in the following way: 
\begin{subequations}
\begin{equation}\label{intrinsic}
\vec r_i  =  \vec x_i- \vec x_x
\end{equation}
\begin{equation}\label{rcm}
\vec R    =  {\frac{m_q(\vec x_1+\vec x_2+\vec x_3) +m_x \vec x_x} {m_t} } 
\end{equation}
\end{subequations}
where, for convenience, we have also introduced the total mass of the constituents:
\begin{equation}
 m_t= 3m_q+ m_x~.
\end{equation}
The previous eqs.  (\ref{intrinsic}) and  (\ref{rcm}) can be inverted, giving:
\begin{subequations}
\begin{equation}\label{xiinv}
 \vec x_i =\vec R + {\frac{ (m_x+ 2 m_q)\vec r_i -m_q (\vec r_j + \vec r_k)} {m_t}}~ 
 , i\neq j\neq k 
\end{equation}
\begin{equation}\label{xxinv}
 \vec x_x=  \vec R -{\frac{   m_q(\vec r_1 +\vec r_2 +\vec r_3) } {m_t}} ~.
\end{equation}
\end{subequations}

\noindent
From the previous eqs. (\ref{xiinv}), (\ref{xxinv}),
we a obtain  the intrinsic momenta $\vec p_i$, conjugate to $\vec r_i$, 
and the total momentum $\vec P$, conjugate to $\vec R$,
in the following form:

\begin{subequations}
\begin{equation}\label{pi}
 \vec p_i = {\frac{ (m_x+ 2 m_q)\vec k_i -m_q (\vec k_j + \vec k_k + \vec k_x)} {m_t}}~ 
 , i\neq j\neq k 
\end{equation}
\begin{equation}\label{ptot}
 \vec P= \vec k_1+ \vec k_2 +\vec k_3 +\vec k_x ~.
\end{equation}
\end{subequations}
Finally, inverting the previous equations, one has:

\begin{subequations}
\begin{equation}\label{pgeni_f_int}
\vec k_i  =  \vec p_i+ {\frac {m_q} {m_t}} \vec P
\end{equation}
\begin{equation}\label{pgenx_f_int}
\vec k_x  =  -(\vec p_1 +\vec p_2 +\vec p_3) + {\frac {m_x} {m_t}} \vec P.
\end{equation}
\end{subequations}
\subsection{The total Hamiltonian}\label{tot_ham}
In the following we shall always  work
in the CM frame of the baryon, where $\vec P=0$.
The Hamiltonian of the model (whose eigenvalues give the baryonic mass spectra)    can be schematically written in the following form:

\begin{equation} \label{hgen}
 H= H_d+ H_{s t} + H_x
\end{equation}
where $H_d$, $H_{s t}$ and $H_x$ respectively represent the  Dirac quark Hamiltonian, 
the spin and isospin dependent Hamiltonian and the kinetic contribution of the $x$-particle. 
\subsection{ The Dirac term}\label{dirac_term}
The Dirac quark term is:
\begin{equation} \label{hd}
 H_d=\sum_{i=1}^3 h(\vec p_i,\vec r_i)
\end{equation}
that represents a sum of three single-particle operators, related to each quark.
Note that, as given by eq. (\ref{pgeni_f_int}), the $\vec p_i$ represent the quark momenta
in the CM; the $\vec r_i$ are the corresponding conjugate coordinates.

\noindent
The single-quark Hamiltonian operator has the form:

\begin{equation} \label{h_i}
  h(\vec p_i,\vec r_i)= \vec \alpha_i \cdot \vec p_i +\beta_i m_q +\omega_i U(r_i)~.
\end{equation}  
The properties of this  single particle Hamiltonian and its solutions  are  studied in detail  in app. \ref{app_dir}.

\noindent
For the specific model, we take the interaction $U(r_i)$ in the form:
\begin{equation}\label{ugen}
 U(r_i)={\frac 1 2} k r_i^2 +U^{(1)}(r_i)
\end{equation}
where the first term represents the confining harmonic oscillator interaction that will be analyzed in   app. \ref{app_ho}.
The second term $U^{(1)}(r_i)$ is taken phenomenologically in the form:


\begin{subequations}
\begin{equation}\label{u1}
\begin{array}{rcl}
& U^{(1)}(r_i)= - {\frac {\tau_c} {r_i}} \left[1-\exp\left( -{\frac {r_i} {ṛ_c }} \right) \right] \\
&  + \Delta_{i} \cdot \left[  \lambda  r_i -   {\frac  {\tau_g}  {r_i} }   \exp(-( {\frac {r_i} {r_g} }) ^2)  \right]~,
\end{array}
\end{equation}
\text{with:}
\begin{equation}
\begin{array}{rcl}
&\Delta_i=1  \text{ for $i=3$ \textit{and} excited states, } \\
&\Delta_i=0  \text{ otherwise.}~~~~~~~~~~~~~~~~~~~~~~~~~
\end{array}
\end{equation}
\end{subequations}
The contribution of the first line represents a regularized Coulombic interaction, 
where $\tau_c$  and $r_c$  are  the effective coupling constant and the regularization radius, respectively.
The interaction of the second line, due to the factor $\Delta_i$, is nonvanishing only for  the quark 3 when it is
in an excited state.
Moreover, the first term represents a linear confining term, besides the harmonic oscillator interaction of eq. (\ref{ugen});
the second term is a short range interaction that has been introduced to reproduce in detail the energy levels 
of the spectra. Due to its short range, it is more effective for the states with $L=l_3=0$.
We recall that also in QDMs, see for example \cite{col_last},  a special term, denoted $M_c(q,r)$, was introduced  
for  the states with $L=0$. 
The coupling costant of the short range interaction has been taken as $\tau_g= \tau_c$ without introducing a new parameter;
finally, the constant $r_g$ represents the radius of the short range interaction.

\vskip 0.5 truecm

\subsection{ The spin-isospin dependent term of the Hamiltonian}\label{hss}
The spin-isospin dependent interaction, that is mainly required to reproduce the spin splittings of the spectra,
is introduced in a phenomenological way,
with a spatial factor that only depends on $\vec r_i$, that is the single quark coordinate.
In this way, we try to simulate the quark-quark  (residual) interaction whose effects cannot be reproduced by the potentials of
eqs. (\ref{ugen}) and (\ref{u1}).

\noindent
The present interaction term, beyond the standard spin-spin and isospin-isospin operators, 
also  depends on $S_c$, $S$, $T$ and $l_i$; the last quantity is the angular momentum quantum number  of the $i$-th quark.
Its expression is inspired by   analogous terms of the QDMs.

\noindent
For clarity we introduce the following  spin-spin operators: 

\begin{subequations}
 \begin{equation}\label{ssc}
  {\cal S}_1={\cal S}_2= {\frac 1 2} \left[ (\vec s_1+\vec s_3)\cdot \vec s_2+ (\vec s_2+\vec s_3)\cdot \vec s_1 \right]
 \end{equation}
 for the interaction of the quarks of the core, and
 \begin{equation}\label{ss3}
 {\cal S}_3= (\vec s_1+\vec s_2) \cdot\vec s_3
 \end{equation}
 for the intearaction of the quark $3$.
\end{subequations}

\noindent
These operators, by definition, are symmetric with respect to interchange of the quarks 1 and 2
according to the general properties of the model.
We also introduce, by replacing $\vec s_i$ with $\vec t_i$ in eqs. (\ref{ssc}) and (\ref{ss3}), the isospin operators ${\cal T}_i$.

\noindent
With those definitions, the spin-isospin dependent Hamiltonian takes the form:

\begin{subequations}
\begin{equation}\label{hstia}
\begin{array}{rcl}
& H^{st}_i=  e^{-\sigma r_i} (-1)^{l_i+1}\{(1 -\Delta_i ) \left[{\cal S}_iA_S + {\cal T}_i A_T  +  {\cal S}_i   {\cal T}_i A_{ST}    \right] ~\\ 
&   ~~~~~~~~~~~~~ + [1+ (-1)^{l_i+1}] \Delta_i [{\cal S}_i \bar A_S + {\cal T}_i  \bar A_T  +  {\cal S}_i   {\cal T}_i\bar A_{ST}    ~                                                          \\
&  +B_S S(S+1) +B_{Sc} S_c(S_c+1) +B_T T(T+1)   ] \} \\
\end{array}
\end{equation}
\text{and finally:}
\begin{equation}\label{hstib}
 H_{st}= \sum_{i=1}^3  H^{st}_i~.
\end{equation}
\end{subequations}
We note that the term proportional to $(1 -\Delta_i )$, due to \textit{this} factor, is active for the quarks of the core $1$, $2$;
for the quark $3$, it gives a  nonvanishing contribution only in the ground states.
On the other hand, the term proportional to $\Delta_i $ gives a contribution for the quark  $3$, only in the excited states.
The matrix elements of the spin  and isospin operators are easily calculated for each state of the model allowing to
determine the total contribution of eq. (\ref{hstib}).

\noindent
We have taken $\bar A_T =\bar A_S$, $ \bar A_{ST} = A_{ST} $ to reduce the number of free parameters of the model
without worsening the reproduction of the experimental spectra.

\subsection{ The kinetic Hamiltonian of the $x$-particle} \label{k_h_x}
The $x$-particle is assumed to be a scalar particle.
In consequence, its kinetic energy is written in the form:
\begin{equation}\label{k_x}
 H_x= \sqrt{m_x^2+(\vec p_1 +\vec p_2 +\vec p_3)^2}
\end{equation}
where  we have used eq. (\ref{pgenx_f_int}) for the momentum $\vec k_x$ of the $x$-particle, in the CM;
furthermore, the product of three Dirac identity operators is understood.

\noindent
If the mean value of the quark momenta is smaller than $m_x$, the standard nonrelativistic expansion
can be performed:
\begin{equation}\label{k_x_nr}
 H_x \simeq m_x+  {\frac {1} {2 m_x}} \cdot (\vec p_1 +\vec p_2 +\vec p_3)^2 ~.
\end{equation}
Note that the products of the momenta of different quarks, 
that is $\vec p_i \cdot \vec p_j $ with $i\neq j$, give vanishing matrix elements
with the wave functions of the model.
For this reason, the nonvanishing matrix elements of  $H_x$ of eq. (\ref{k_x_nr}) are proportional to the
squared quark momenta $\vec p_i^2$,
that are single particle operators.
Their contributions are calculated perturbatively and added to the total energy of the resonances.
We finally note that the $x$-particle mass, $m_x$, as shown by eq. (\ref{k_x_nr}),  represents, at the same time, the zero point energy
of the spectrum.

\section{ The magnetic moment of the nucleon}\label{mag_mom}
In this section we study the static magnetic properties of the $N(939)$.
The interaction of the system with an external electromagnetic three-vector  field  $\vec A$ is introduced by means of the  minimal substitution
on the quark momenta; the $x$-particle, being electrically neutral does not contribute.
The minimal substitution has the standard  form:
$\vec k_i \rightarrow \vec k_i -e_i {\vec A}({\vec x_i}) $
where $\vec x_i$ and $\vec k_i$ are the quark coordinate and momenta in a generic frame and $e_i$ represents the electric charge of the i-th quark.
However, considering that in our model $m_q << m_x$,  (see the numerical values of the parameters in table  \ref{res_par}  ) by means of eq. (\ref{pi}),
one finds that the minimal substitution can be performed directly on the intrinsic quark momenta $\vec p_i$ that appear in
the Hamiltonian of the model.
For the same reason, by means of eq. (\ref{xiinv}), with $\vec R =0$, one can also approximate the generic frame coordinates with the relative ones, 
that is $\vec x_i \simeq \vec r_i$.

\noindent
In more detail, we make the minimal substitution in the Dirac term of the Hamiltonian, given by eq.(\ref{h_i}); 
the spin-isospin term  $H_{st}$ of eq.(\ref{hstia}) does not contain the quark momenta and, in consequence, gives no contribution;
the kinetic operator of the $x$-particle, does contain the quark momenta and, in principle,  could only give a contribution
to the \textit{orbital terms} of the magnetic moment of the nucleon.
However we recall that for the  $N(939)$, all the quark orbital angular momenta are vanishing;
in consequence, only the \textit{spin terms} derived from $H_d$, give a contribution to the nucleon magnetic moment.

\noindent
For the reasons discussed above, we can take the total magnetic dipole of the system as the sum of the single-quark contributions
and make use of the results obtained in app. \ref{app_magdip}, in particular the development of eq. (\ref{H_s}) and the final result
of eq. (\ref{gd0}).
In this way we can write the total magnetic dipole operator in the spin-isospin space, in the following form:
\begin{equation}\label{mudip_tot}
\vec \mu= \sum_{i=1}^3 e_i G^{(d)}_{0, i} \vec \sigma_i \rightarrow 3 e_3 G^{(d)}_{0} \vec \sigma_3
\end{equation}
where, in the last expression on the right, we have taken into account the (anti)symmetry of the nucleon wave function; we have also dropped the quark index
$i$  in $G^{(d)}_{0}$ recalling that the three quarks have the same spatial wave function, that is ${\frac {1} {\sqrt{4\pi}} } R_{0,0}(r_i)$,
as explained in app. \ref{app_wf}.

\noindent
The magnetic moments of the nucleon  are obtained calculating the mean values of $\mu_z$ of eq. (\ref{mudip_tot})
with the spin and isospin factor of the wave function of eq. (\ref{Phi_N_b}), taking $M_J=M_s=1/2$
and $M_T=\pm 1/2$ for the proton and neutron, respectively.

\noindent
The calculation is performed in the same way as in the CQMs, replacing ${\frac {1} {2m}}$ with $G^{(d)}_{0}$.
The results, in nuclear magneton units, are:
\begin{subequations}\label{mufinal}
 \begin{equation}
  \mu_p=2 M_p G^{(d)}_{0}
 \end{equation}
 for the proton, and
\begin{equation}
 \mu_n=-{\frac 4 3} M_p G^{(d)}_{0}
\end{equation}
for the neutron,
\end{subequations}
where $M_p$ represents the proton mass.

\noindent
Note that  the ratio of the proton and neutron
magnetic moments  does not depend on the value of $  G^{(d)}_{0}$ and is, in any case, $\mu_p/ \mu_n=- 3/2$.
The numerical results, obtained with the solutions of the Hamiltonian wave equation,  for $G^{(d)}_{0}$,  $\mu_p$ and $\mu_n$, are given in table \ref{res_mu}.

\section{ The results for the spectra and the nucleon magnetic moments}\label{results}
The results of our Dirac shell-core model calculation, compared with the experimental data \cite{pdg},
are shown in table \ref{tab_n_res} and table \ref{tab_d_res}, for the
$N$ and $\Delta$ states, respectively.
The theoretical results are  obtained with two sets of slightly different parameters, namely $(a)$ and $(b)$, given in table \ref{res_par}.
The relatively high number of parameters is related to the phenomenological character of the model in which different effects of the  interaction
are parametrized by means of the potential terms introduced above.
In the present work  we have, totally, $14$ free parameters, considering that the quark mass $m_q=3.5~ MeV$ 
is obtained from  QCD extimations, as explained in sect. \ref{general};
for a comparison, in  the QDM  of ref. \cite{col_last}, $15$ parameters were used to 
reproduce the $N$ and $\Delta$ spectra.
The value of $m_x$ in the present work (see table \ref{res_par}) is greater  but of the same order of magnitude as
 the zero point energy used in ref. \cite{col_last}, that was $E_0= 826~ MeV$.\\
All the experimental data of tables \ref{tab_n_res} and \ref{tab_d_res} have been taken into account
to determine (by means of a complex fit procedure) the free parameters of the model. 
We point out that the quantum number \textit{assignations} of table \ref{tab:nfirststates} and table \ref{tab:deltafirststates}
represent a crucial element to perform  the whole process.
Moreover, for the degenerate multiplets (with respect to $J$) the central values of the corresponding experimental mass data
have been used.\\
Our model reproduces all the $3^*$ and $4^*$ resonances up to 2 GeV using the states listed in table \ref{tab:nfirststates}.
For the $N(1880) {\frac 1 2}^+$  only,  we have used $n_r=2$. 

\noindent
The experimental masses are reproduced with acceptable accuracy. 
A slight improvement is obtained with respect to the QDM of ref.\cite{col_last}.
In general, some discrepancies with the experimental data are found in the degenerate multiplets, given that the spin-orbit interaction
has not been included in the model.

\noindent
Analyzing  the $N$ resonances of table \ref{tab_n_res}, we note that the theoretical mass for the $N(1900) {\frac 3 2}^+$ is lower than
the experimental data; a better extimation is given by the parameter set $(b)$.
In any case an improvement is obtained with respect to ref. \cite{col_last}, where the result of the calculation was $1780 ~ MeV$.
We also note that, for this resonance, the experimental mass interval passed from $1870 - 1930 ~MeV$ 
of the previous \textit{Particle Data Group} \cite{pdg_old}, to the actual value of $1890 - 1950 ~ MeV$.

\noindent
Our model  predicts, with a theoretical mass of $1970~ MeV$ (set $(a)$) and $1983 ~MeV$ (set $(b)$),  a state   $N {\frac 5 2}^+$. 
This state is associated to the resonance $N(2000) {\frac 5 2}^+$, that is a $2^*$ resonance.
We also have a $N {\frac 3 2}^+$, degenerate with the former. This state is tentatively assigned to the  $N(2040) {\frac 3 2}^+$ resonance,
that is a $1^*$ resonance.
Below $2 ~GeV$  no other missing resonances are predicted.
Finally, the model predicts a positive parity multiplet , with $ {\frac 1 2} \geq J \geq  {\frac 7 2}$ at $ 2090 ~MeV$ (set $(a)$) and $2104 ~MeV$ (set $(b)$). 
Experimentally, only the $N(2100)  {\frac 1 2}^+$ is observed.
%
The $N(2040) {\frac 3 2}^+$  and the $N(2000) {\frac 5 2}^+$ are the only $1^*$ and $2^*$  $N$ resonances reported in table \ref{tab_n_res}.

We now analyze the $\Delta$ resonances of table \ref{tab_d_res}. \\
We note that the $\Delta(1600) {\frac 3 2}^+$ is not reproduced accurately
by our model, in particular by set $(b)$. 
We note that, for this resonance, the experimental mass interval passed from $1500 - 1700 ~ MeV$ 
of the previous \textit{Particle Data Group} \cite{pdg_old}, 
to the actual value of $1500 - 1640 ~ MeV$.
The model predicts, besides the $3^*$ and $4^*$ $\Delta$ resonances up to $2~ GeV$, a state  $\Delta {\frac 1 2}^+$ with a theoretical mass
of $ 1759 ~ MeV$ (set $(a)$) and $ 1779 ~ MeV$ (set $(b)$). This state is associated to the $\Delta(1750) {\frac 1 2}^+$, that is a $1^*$ resonance.

\noindent
Considering the triplet with $L=1$ and $S=3/2$ and $1/2 \geq J \geq 5/2$, the member with $J= 3/2$ is associated to the $\Delta(1940) {\frac 3 2}^-$, 
 $2^*$ resonance;
for this state our model predicts a mass value of $1902 ~MeV$ and $1903 ~MeV$, with set $(a)$ and set $(b)$, respectively.

\noindent
Finally, our model predicts a doublet with $L=2$ and $S=1/2$ at $2030 ~MeV$ (set $(a)$) and $2043 ~ MeV $ (set$(b)$).
The member of the doublet with $J= 5/2$ is associated to the  $\Delta(2000) {\frac 5 2}^+$, $2^*$ resonance.
The other member of the doublet, with $J= 3/2$ is not observed experimentally.

\noindent
In table \ref{tab_d_res} we have reported only the three $1^*$ and $2^*$ resonances mentioned above.

In table \ref{res_mu} we give the results for the factor $G^{(d)}_{0}$ and the magnetic moments of the nucleon.
As in CQMs, the results favourably compare with the experimental data.

\section{Conclusions and outlook}\label{conclusion}
In this work we have developed a Dirac quark shell model to study the baryonic spectra.
The experimental data are well reproduced taking  into account only the one-quark excitations.\\
With respect to CQMs, our model does not introduce any missing resonance  up to $2000~MeV$.
With respect to the QDMs, we obtain baryonic spectra of the same quality.
However, our model presents some relevant improvements considering its theoretical consistency.  
Namely, the diquark is replaced by the two unexcited quarks of the core, 
without the necessity of introducing a specific \textit{freezing} hypothesis.
Moreover, the quark not belonging to the core, having different physical \textit{effective} properties,
is not identical to the two quarks of the core and does not require wave fuction antisymmetrization.
Finally, the quark wave function is completely relativistic.
The use of the Dirac equation with equal scalar and vector potentials (spin symmetry case),
avoids the spin-orbit splittings and, moreover, allows to take
the QCD value for the quark mass,
without the necessity of introducing a \textit{dressing} mechanism for this parameter.
The same structure of the adopted Dirac equation, also gives, with a \textit{small} quark mass,
the correct value for the nucleon magnetic moment.


\noindent
Further investigation is needed to understand, at a more fundamental level, the reason why
only  one-quark excitations reproduce the baryonic spectra.
Finally, the spin-orbit and tensorial interactions should be introduced and
the quark-quark residual interaction should be also studied to construct a a complete model for the baryonic spectroscopy.


\begin{table*}

\caption{Comparison between the experimental values \cite{pdg}  of the $N$ resonance
masses up to 2 GeV and the results of the model (all mass values are expressed in MeV). 
Two resonances predicted by the model, with  experimental masses above 2 GeV, are shown at the bottom of the table.
The quantum numbers $J^P$, $n_r$, $L$, $S_c$ and $S$  have been introduced in sect. \ref{general};
they represent the total angular momentum and parity, the radial excitation, the total orbital angular momentum,
the core spin and the total spin, respectively.
The states with $S={\frac 3 2}$  have necessarily $S_c=1$.}
\begin{center}
\begin{tabular}{cccccccccc}
\hline
\hline \\
Resonance & Status & $M^{exp.}$        &     $J^P$        & $n_r$    & $L$ & $S_c$  & $S $          & $M_a^{calc.}$  & $M_b^{calc.}$      \\
          &        & (MeV)             &                  &          &     &        &               &  (MeV)         & (MeV)               \\ \\
\hline \\
$N(939)$  & ****   &  939              &  $\frac{1}{2}^+$ & $0$      & $0$ & $(0,1)$& $\frac{1}{2}$ & 939            &  938                \\
$N(1440)$ & ****   &  1410 - 1470      &  $\frac{1}{2}^+$ & $1$      & $0$ & $0$    & $\frac{1}{2}$ & 1429           & 1446                \\
$N(1520)$ & ****   &  1510 - 1520      &  $\frac{3}{2}^-$ & $0$      & $1$ & $0$    & $\frac{1}{2}$ & 1510           & 1517                 \\
$N(1535)$ & ****   &  1515 - 1545      &  $\frac{1}{2}^-$ & $0$      & $1$ & $0$    & $\frac{1}{2}$ & 1510           & 1517                 \\
$N(1650)$ & ****   &  1635 - 1665      &  $\frac{1}{2}^-$ & $0$      & $1$ & $1$    & $\frac{3}{2}$ & 1672           & 1677                  \\
$N(1675)$ & ****   &  1665 - 1680      &  $\frac{5}{2}^-$ & $0$      & $1$ & $1$    & $\frac{3}{2}$ & 1672           & 1677                 \\
$N(1680)$ & ****   &  1680 - 1690      &  $\frac{5}{2}^+$ & $0$      & $2$ & $0$    & $\frac{1}{2}$ & 1698           & 1710                 \\
$N(1700)$ & ***    &  1650 - 1800      &  $\frac{3}{2}^-$ & $0$      & $1$ & $1$    & $\frac{3}{2}$ & 1672           & 1677                  \\
$N(1710)$ & ****   &  1680 - 1740      &  $\frac{1}{2}^+$ & $1$      & $0$ & $1$    & $\frac{1}{2}$ & 1700           & 1719                  \\
$N(1720)$ & ****   &  1680 - 1750      &  $\frac{3}{2}^+$ & $0$      & $2$ & $0$    & $\frac{1}{2}$ & 1699           & 1710                   \\
$N(1875)$ & ***    &  1850 - 1920      &  $\frac{3}{2}^-$ & $0$      & $1$ & $1$    & $\frac{1}{2}$ & 1871           & 1882                  \\
$N(1880)$ & ***    &  1830 - 1930      &  $\frac{1}{2}^+$ & $2$      & $0$ & $0$    & $\frac{1}{2}$ & 1847           & 1865                  \\
$N(1895)$ & ****   &  1870 - 1920      &  $\frac{1}{2}^-$ & $0$      & $1$ & $1$    & $\frac{1}{2}$ & 1871           & 1882                  \\
$N(1900)$ & ****   &  1890 - 1950      &  $\frac{3}{2}^+$ & $1$      & $0$ & $1$    & $\frac{3}{2}$ & 1820           & 1840                  \\
$N(2000)$ & **     &  1950 - 2150      &  $\frac{5}{2}^+$ & $0$      & $2$ & $1$    & $\frac{1}{2}$ & 1970           & 1983               \\  \\
\hline \\
$N(2040)$ & *      &  2010 - 2070      &  $\frac{3}{2}^+$ & $0$      & $2$ & $1$    & $\frac{1}{2}$ & 1970           & 1983               \\
$N(2100)$ & ***    &  2050 - 2150      &  $\frac{1}{2}^+$ & $0$      & $2$ & $1$    & $\frac{3}{2}$ & 2090           & 2104                \\  \\
\hline
\hline
\end{tabular}
\end{center}
\label{tab_n_res}
\end{table*}


\begin{table*}

\caption{Comparison between the experimental values \cite{pdg}  of the $\Delta$ resonance
masses up to 2 GeV and the results of the model.
At the bottom of the table we also show a resonance, predicted by the model, whose real experimental mass is greater than 2 GeV.  
The units for the masses and the quantum numbers 
are as in table  \ref{tab_n_res}.
 For all the resonances one  has necessarily $S_c=1$, that has been omitted in the table.}
\begin{center}
\begin{tabular}{ccccccccc}
\hline
\hline \\
Resonance      & Status & $M^{exp.}$     & $J^P$             & $n_r$    & $L$   & $S $           & $M_a^{calc.}$  &   $M_b^{calc.}$  \\
               &        & (MeV)          &                   &          &       &                &  (MeV)         &    (MeV)\\ \\
\hline \\
$\Delta(1232)$ & ****  &  1230 - 1234   & $\frac{3}{2}^+$    & $0$      & $0$   & $\frac{3}{2}$  &  1230          &   1230  \\
$\Delta(1600)$ & ****  &  1500 - 1640   & $\frac{3}{2}^+$    & $1$      & $0$   & $\frac{3}{2}$  &  1678          &   1698  \\
$\Delta(1620)$ & ****  &  1590 - 1630   & $\frac{1}{2}^-$    & $0$      & $1$   & $\frac{1}{2}$  &  1687          &   1672  \\
$\Delta(1700)$ & ****  &  1690 - 1730   & $\frac{3}{2}^-$    & $0$      & $1$   & $\frac{1}{2}$  &  1687          &   1672 \\
$\Delta(1750)$ & *     &  1680 - 1782   & $\frac{1}{2}^+$    & $1$      & $1$   & $\frac{1}{2}$  &  1759          &   1779  \\
$\Delta(1900)$ & ***   &  1840 - 1920   & $\frac{1}{2}^-$    & $0$      & $1$   & $\frac{3}{2}$  &  1902          &   1903 \\
$\Delta(1905)$ & ****  &  1855 - 1910   & $\frac{5}{2}^+$    & $0$      & $2$   & $\frac{3}{2}$  &  1949          &   1962  \\
$\Delta(1910)$ & ****  &  1850 - 1950   & $\frac{1}{2}^+$    & $0$      & $2$   & $\frac{3}{2}$  &  1949          &   1962  \\
$\Delta(1920)$ & ****  &  1870 - 1970   & $\frac{3}{2}^+$    & $0$      & $2$   & $\frac{3}{2}$  &  1949          &   1962  \\
$\Delta(1930)$ & ***   &  1900 - 2000   & $\frac{5}{2}^-$    & $0$      & $1$   & $\frac{3}{2}$  &  1902          &   1903  \\
$\Delta(1940)$ & **    &  1940 - 2060   & $\frac{3}{2}^-$    & $0$      & $1$   & $\frac{3}{2}$  &  1902          &   1903 \\ 
$\Delta(1950)$ & ****  &  1915 - 1950   & $\frac{7}{2}^+$    & $0$      & $2$   & $\frac{3}{2}$  &  1949          &   1962  \\  \\
\hline\\
$\Delta(2000)$ &**     &  2075 - 2325   & $\frac{5}{2}^+$    & $0$      & $2$   & $\frac{1}{2}$  &  2030          &   2043   \\ \\
\hline
\hline
\end{tabular}
\end{center}
\label{tab_d_res}
%
\end{table*}

\begin{table}[h]  
\caption{Values of the model parameters.}
\begin{center}
\begin{tabular}{llll}
\hline
\hline \\
         &   set $(a)$            &  set  $(b)$  &  units     \\ 
\hline \\        
 $m_q$   &  $~3.5$                & $~3.5$       &  MeV   \\
 $m_x$   &  $~1.574$              & $~1.570$     &  GeV   \\
 $k  $   &  $~0.1611$             & $~0.1627$    &  GeV $\cdot$ fm $^{-2}$  \\
 $\tau_c$&  $~4.292 $             & $~4.283 $    &         \\
 $r_c$   &  $~0.6695$             & $~0.6692$    &  fm     \\
 $r_g$   &  $~0.3322$             & $~0.3478$    &  fm     \\
 $\lambda$& $~0.2572$             & $~0.2442$    &  GeV $\cdot$ fm $^{-1}$  \\
 $\sigma $& $~2.326 $             & $~2.325 $    &  fm $^{-1}$  \\
 $A_S    $& $-0.3293$             & $-0.3326$    &  GeV    \\
 $A_T    $& $~52.01 $             & $~50.62 $    &  MeV    \\
 $A_{ST} $& $~1.568 $             & $~1.569 $    &  GeV    \\
 $\bar A_S$&$~0.2971 $            & $~0.2952$    &  GeV    \\
 $ B_S   $ &$~75.15 $             & $~74.38 $    &  MeV    \\
 $ B_{Sc}$ &$-0.1845$             & $-0.1845$    &  GeV    \\
 $ B_T   $ &$~0.1593$             & $~0.1562$    &  GeV    \\
 \\
\hline
\hline
\end{tabular}
\end{center}

\label{res_par}
\end{table}
\begin{table}[h]  
\caption{Results for $G^{(d)}_{0}$  and proton and neutron magnetic moments.}
\begin{center}
\begin{tabular}{lllll}
\hline
\hline \\
               &   set $(a)$             &  set  $(b)$  & exp.        &  units     \\ 
\hline \\        
$G^{(d)}_{0}$  & $~ 1.506   $            & $~ 1.502   $ &             &  GeV $^{-2}$ \\
$\mu_p      $  & $~ 2.826   $            & $~ 2.819   $ & $~2.793$    &  n.m.u.      \\
$\mu_n      $  & $- 1.884   $            & $- 1.879   $ & $-1.913$    &  n.m.u.      \\ 
 \\
\hline
\hline
\end{tabular}
\end{center}

\label{res_mu}
\end{table}

\newpage
\vskip 6.0 truecm

\appendix

\vskip 5.0 truecm

\section{ Dirac equation with spin symmetry}\label{app_dir}

The Hamiltonian of eq.(\ref{h_i}) 
represents the sum of three \textit{independent} Hamiltonian operators (with spin symmetry) 
for the quarks of the baryon.
We now discuss some general properties  of a \textit{single particle} Hamiltonian in the case of spin symmetry.
For simplicity, in this discussion, we shall drop the quark index $i$
and also  put $m_q=m$.

\noindent
The Hamiltonian operator has the form:

\begin{equation} \label{h}
  h(\vec p,\vec r)= \vec \alpha \cdot \vec p +\beta m  +\omega U(r)
\end{equation}  
where we have introduced the standard Dirac matrices $ \vec \alpha= \gamma^0 \vec\gamma $, $\beta= \gamma ^0$,
in the standard representation,
and the following projection operator:
\begin{equation}\label{omega}
 \omega={\frac 1 2}   (1+ \beta)~.
\end{equation}
%
Preliminarily, we split the Dirac spinor into two upper and two lower components, that is:
\begin{equation}
 \psi(\vec r) = 
\begin{pmatrix}
\hat \varphi(\vec r)\\
\hat \eta(\vec r)
\end{pmatrix} ~.
\end{equation}
By applying the $\omega$ projection operator to the Dirac spinor, one immediately finds:
\begin{equation}\label{project}
 \omega ~\psi(\vec r)= 
 \begin{pmatrix}
\hat \varphi(\vec r)\\
0 
\end{pmatrix} ~.
\end{equation}
We note that the projection operator $\omega$ annihilates the lower components of a Dirac spinor.
Going back to eq. (\ref{h}), we also note that the
interaction  operator $\omega U(r)$  contains 
a time component of a vector interaction and a scalar interaction,   
respectively given by the first and second term of $\omega$, as shown by eq. (\ref{omega}).
These two terms have the same spatial dependence:
\begin{equation}
 V_s(r)=V^0_v(r)= {\frac 1 2} U(r)~.
\end{equation}
The eigenvalue equation   
corresponding to the Hamiltonian of eq. (\ref{h}) is:
\begin{equation}\label{eig_h}
  h(\vec p,\vec r)\psi(\vec r)= E \psi(\vec r)~.
\end{equation}
Taking into account eq. (\ref{project}),
we can rewrite
eq. (\ref{eig_h}) as two coupled equations, in the following form:

\begin{subequations}
\begin{equation}\label{eq:indea}
\vec{\sigma}\cdot\vec p \hat \eta(\vec r)  +(m+U(r))\hat\varphi(\vec r)   =E\hat \varphi (\vec r)
\end{equation}
\begin{equation}\label{eq:indeb}
\vec{\sigma}\cdot\vec p \hat \varphi(\vec r)   -m\hat \eta (\vec r)  =E\hat \eta(\vec r)
\end{equation}
\end{subequations}
where  $\vec \sigma$ represents the vector of the three Pauli matrices.
The previous equations
can be solved expressing the lower components $\hat \eta(\vec r)$  of eq. (\ref{eq:indeb}) by means of the upper ones $\hat \varphi(\vec r)$;
replacing the result in eq. (\ref{eq:indea}), without approximations, 
one obtains
a Schr\"odinger-like, \textit{energy-dependent}, equation 
in the form \cite{mdes_edep}:
\begin{equation}\label{equpper}
\left( {\frac {\vec p^2} {E+m}} + U(r)+m \right)\hat \varphi(\vec r)= E \hat \varphi(\vec r) 
\end{equation}
where we require $E \neq -m$. See also, in the following, eq. (\ref{equpper_bis}).
Note that, in eq. (\ref{equpper}), the spin does not appear explicitly; 
in consequence, \textit{no spin-orbit effect} is introduced and the  spin dependence can be completely factorized.
We have:
\begin{equation}\label{phi_with_chi}
 \hat \varphi(\vec r)= \varphi(\vec r)   \chi_{m_s},
\end{equation}
where $\varphi(\vec r) $ is a one-component function  and $ \chi_{m_s}$ is a standard Pauli spinor 
corresponding to the state $|1/2, m_s>$.

\noindent
The lower components of the Dirac spinor  $ \hat \eta (\vec r)$  can be obtained straightforwardly from eq. (\ref{eq:indeb});
in this way the complete four-component Dirac spinor takes the form:
\begin{subequations}
\begin{equation}\label{psidira}
\psi(\vec r) = N 
\begin{pmatrix}
1\\
\frac{\vec{p}   \cdot\vec{\sigma} }{E +m}
\end{pmatrix}
 \varphi(\vec r)   \chi_{m_s} =
\end{equation}
\begin{equation}\label{psidirb}
 ~=D(\vec p\cdot\vec \sigma; E, m) \varphi(\vec r)   \chi_{m_s}~.
\end{equation}
\end{subequations}
In eq. (\ref{psidira}) we have introduced the normalization constant $N$, that will be determined in the following.
Eq. (\ref{psidirb}) synthetically defines  the operator $ D(\vec p \cdot\vec \sigma; E, m) $ that constructs the  four components Dirac spinor
when it is applied to the corresponding two component spinor. This operator will be used when studying the complete wave function 
of the baryonic system in app. \ref{app_wf}.

\noindent
To calculate the normalization constant $N$, we preliminarily introduce the normalization integral:
\begin{subequations}
\begin{equation}\label{inorm}
 I= < \varphi| 1+  {\frac {\vec p ^2} {(E+m)^2} } | \varphi >~,
\end{equation}
\text{so that:}
\begin{equation}\label{norm}
N=I^{-1/2}~.
\end{equation}
\end{subequations}
We recall that, when solving eq. (\ref{equpper}), we shall also diagonalize the orbital angular momentum. 
In consequence, we shall introduce 
the indices $n,~l, m_l$ that respectively denote
the number of nodes in the radial wave function and the quantum numbers of the orbital angular momentum.
In particular, for the upper component wave function, we have:
\begin{equation}\label{phi_nlml}
\varphi(\vec r) =\varphi_{n, l, m_l}(\vec r)=R_{n,l}(r) Y_{l, m_l}(\hat r)~.
\end{equation}
Also note that eq. (\ref{equpper}), due to its energy-dependence, does \textit{not} represent an eigenvalue equation for a Hermitean operator,
for this reason its solutions are \textit{not} orthogonal with respect to index  $n$:
%
$< \varphi_{n',l, m_l}| \varphi_{n,l, m_l}> \neq \delta_{n',n}$;
on the contrary, the Dirac spinors of eqs. (\ref{psidira}), (\ref{psidirb}), with the normalization of eq. (\ref{norm}), 
being the eigenstates of the Hermitean Dirac Hamiltonian of eq. (\ref{h_i}),
do satisfy standard orthonormality:

\begin{equation}\label{orthonorm}
<\psi_{n',l', m_l',m_s'}|\psi_{n,l, m_l,m_s} >=\delta_{n',n} \cdot \delta_{l',l} \cdot \delta_{m_l',m_l} \cdot \delta_{m_s',m_s} 
\end{equation}
\section{ The Harmonic Oscillator interaction}\label{app_ho}
For some forms of $U(r)$  eq. (\ref{equpper}) can be solved analytically
by using the results of the ``corresponding'' nonrelativistic equation.
In this section we study in detail the case of a harmonic interaction.

\noindent
\textit{In general}, we introduce for convenience the ``subtracted'' energy $\bar E$:
\begin{equation}\label{ebar}
\bar E= E-m ~.
\end{equation}
In this way eq. (\ref{equpper}) can be written as:
\begin{equation}\label{equpper_bis}
\left( {\frac {\vec p^2} {2( m+ { \frac {\bar E} 2})}} + U(r) \right)\varphi(\vec r)= \bar E  \varphi(\vec r) 
\end{equation}
where we have also discarded the Pauli spinor $\chi_{ms}$, taking into account eq. (\ref{phi_with_chi}).
We note that eq. (\ref{equpper_bis}) can be obtained from the nonrelativistic Schr\"odinger equation
by replacing:

\begin{subequations}
\begin{equation} \label{subse}
 E^{(nr)} \rightarrow \bar E
\end{equation}
\begin{equation}\label{subsm}
 m \rightarrow m + {\frac {\bar E} {2} }~.
\end{equation}
\end{subequations}
\noindent
\textit{In particular},  we consider a harmonic oscillator   (HO) interaction:
\begin{equation}\label{uho}
 U(r)= {\frac 1 2} k r^2
\end{equation}
%
We recall that in the nonrelativitic case, the energy eigenvalues are:
\begin{equation}\label{enr}
 E^{(nr)}_{n_e}=(n_e+ {\frac 3 2} )  \sqrt{ {\frac {k}  { m   }   } }
\end{equation}
where we have introduced  for convenience the energy quantum number $n_e$ that is related
to the number of nodes $n$ by the standard equation:
\begin{equation}\label{ne}
 n_e=2 n+ l .
\end{equation}
To solve the relativistic problem of eq. (\ref{equpper_bis}), with the interaction of  eq. (\ref{uho}),
we make the replacement of eq. (\ref{subsm}) in eq. (\ref{enr}),    obtaining:
\begin{subequations}
\begin{equation}\label{ebarsol_a}
\bar E= (n_e+ {\frac 3 2} )  \sqrt{ {\frac {k}  { m+  {\frac {\bar E} {2}}      }   } }
\end{equation}
\text{or, equivalently:}
\begin{equation}\label{ebarsol_b}
\bar E=  E^{(nr)}_{n_e} \sqrt{ {\frac {m}  { m+ {\frac {\bar E} {2}}      }   } }~.
\end{equation}
\end{subequations}
This equation  can be transformed into a cubic equation for $\bar E$ and solved analytically, 
finding the energy values $\bar E_{n_e}$. 
By means of eq. (\ref{ebar}) one has $E_{n_e}= \bar E_{n_e} +m$.

%

\noindent
In order to determine the form of the
radial wave functions, we recall that in the nonrelativistic HO case, these functions depend on
%
the dimensional constant $\bar r$,
that is given by the equation:
\begin{equation}\label{rbar_nr}
 \bar r= (mk)^{-1/4}~.
\end{equation}
We also write the nonrelativistic harmonic HO radial wave functions as:

\begin{subequations}
\begin{equation}\label{radial_nr}
 R_{n,l}^{(nr)}(r; \bar r) = (\bar r)^{-3/2} S_{n,l}(x)
\end{equation}
\text{with}
\begin{equation}\label{x_nr}
 x={\frac {r} {\bar r}}~~.
\end{equation}
\end{subequations}
For completeness, we also recall that:
\begin{equation}\label{stand_rad_ho}
 S_{n,l}(x)=\left[ {\frac {2 ( n !)} {\Gamma({n+l+{\frac 3 2} })} }\right]^{\frac 1 2} x^l {\cal L}_n^{l+ {\frac 1 2}}(x^2)\exp(- {\frac 1 2} x^2)
\end{equation}
where ${\cal L}_n^{l+ {\frac 1 2}}(x^2)$ are the generalized Laguerre polynomials.
We can now calculate the
$R_{n,l}(r)$ for our relativistic problem by
 performing the substitution of eq. (\ref{subsm}) for the mass $m$ in eq. (\ref{rbar_nr});
then, by using eq. (\ref{ebarsol_b}), one obtains:
\begin{equation}\label{rbar}
 \bar r_{n_e}= \sqrt{ {\frac  { \bar E_{n_e}}   {E_{n_e}^{(nr)}}  } } \cdot \bar r
\end{equation}
and, in consequence:
\begin{subequations}
\begin{equation}\label{radial}
 R_{n,l}(r;\bar r_{n_e}) = (\bar r_{n_e})^{-3/2} S_{n,l}(x_{n_e})
\end{equation}
\text{with}
\begin{equation}\label{x_n_e}
 x_{n_e}={\frac {r} {\bar r_{n_e}}}~.
\end{equation}
\end{subequations}
Note that, in the relativistic case, the dimensional constant $\bar r_{n_e}$ is energy-dependent.

\noindent
Finally, also the normalization integral of eq. (\ref{inorm}) can be calculated analytically.
From this quantity, one obtains the normalization constant of eq. (\ref{norm}), in the form:
\begin{equation}\label{inorm_ho}
 N_{n_e}=\left[ 1+ (\bar E_{n_e} +2m)^{-3/2} \left({\frac {m} {2}}\right)^{1/2} E_{n_e}^{(nr)} \right] ^{-1/2}~.
\end{equation}
Collecting all the results obtained above, we can write the Dirac orthonormal wave functions for HO interaction in the form:

\begin{equation}\label{psidir_ho}
\psi_{n;l,m_l;m_s}(\vec r) = N_{n_e} 
\begin{pmatrix}
1\\
\frac{\vec{p}   \cdot\vec{\sigma} }{E_{n_e} +m}
\end{pmatrix}
  R_{n,l}(r;\bar r_{n_e}) Y_{l,m_l}(\hat r)   \chi_{m_s}.
\end{equation}
As in eq. (\ref{psidirb}), the operator $D(\vec p\cdot\vec \sigma; E_{n_e}, m)$
can be introduced.

\noindent
A wave function with the same spin-angular quantum numbers  ($l,m_l, m_s$), but with a different radial dependence
can be expanded by means of the eigenfunctions of eq. (\ref{psidir_ho}) in the form:

\begin{equation}\label{expansion}
 \psi^{(g)}_{n';l,m_l;m_s}(\vec r)= \sum_{n=0}^{n_{max}} a^{n'}_{n;l,m_l;m_s} \psi_{n;l,m_l;m_s}(\vec r)
\end{equation}
where the upper index $g$ in the wave function of the \textit{l.h.s.} denotes its general character.

In the present work we use that expansion for the single quark wave functions.
The amplitudes $a^{n'}_{n;l,m_l;m_s}$ are determined by diagonalyzing the interaction operator in the
relativistic HO basis given by the wave functions of eq. (\ref{psidir_ho}).
We point out  that the index $n'$ corresponds to the radial excitation number $n_r$ introduced in eq. (\ref{psi}) of sect. \ref{general}.

\section{ The magnetic dipole operator}\label{app_magdip}
We now study the magnetic dipole operator for the one-particle Dirac equation with spin symmetry.
We recall that
in the case of a free Dirac equation, when the interaction with a magnetic field is introduced, one obtains
the well-known  result for the magnetic dipole operator of a point-like, \textit{free} particle:
\begin{subequations}
\begin{equation}\label{mudirac}
\vec \mu = e ~ G^{(f)} \vec \sigma~,
\end{equation}
with
\begin{equation}
 G^{(f)}={\frac {1} {2m}}~.
\end{equation}
\end{subequations}
In this section we shall derive an analogous expression for the Dirac equation with interaction in the case of spin symmetry.
We start by performing the minimal coupling substitution in eq. (\ref{h}).  
We obtain the standard result for the interaction Hamiltonian with
an external three-vector field $\vec A(\vec r) $, that is:
\begin{equation}\label{hint}
 H_{int} =- e \vec \alpha \cdot \vec A(\vec r) 
\end{equation}
For studying our shell quark model,
we consider  two different Dirac wave functions
$|\psi_a>$, $|\psi_b>$ and calculate the matrix element of the operator of eq. (\ref{hint})
between these wave functions.
We do not include in the matrix element the two component spinors $\chi_{m_{s_{a}}}$,  $\chi_{m_{s_{b}}}$,
in order to highlight, as in eq. (\ref{mudirac}), the dependence of the magnetic dipole  operator on the Pauli matrices $\vec \sigma$.
Furthermore,
we shall use the spatial wave functions  $|\varphi_a>$, $|\varphi_b>$
introduced in eqs. (\ref{phi_with_chi}), (\ref{psidira}) and (\ref{psidirb}).
With standard handlings one obtains:
\begin{equation}\label{hintspit}
 <\psi_b|H_{int}|\psi_a> = <\psi_b|H_o|\psi_a> + <\psi_b|H_s|\psi_a>
\end{equation}
where the first term represents the \textit{orbital contribution}, of the form:
\begin{equation}\label{hint_o}
<\psi_b|H_o|\psi_a>= -eN_bN_a <\varphi_b | {\frac {\vec A \cdot \vec p}  {E_a+m} }+
                                           {\frac {\vec p \cdot \vec A}  {E_b+m} } |\varphi_a>.
\end{equation}
We shall not develop further this term and focus our attention
on the second term that gives the \textit{spin contribution}:
\begin{equation}\label{hint_s}
<\psi_b|H_s|\psi_a>= -ieN_bN_a \vec \sigma \cdot <\varphi_b |  {\frac {\vec A \times \vec p}  {E_a+m} }+
                                           {\frac {\vec p \times \vec A}  {E_b+m} } |\varphi_a>.
\end{equation}
We consider the case of $|\psi_a>=|\psi_b>=|\psi>$,  and, in consequence, $E_a=E_b=E$, etc..
Furthermore, we take a uniform magnetic field $\vec B$, given by
$\vec B= \vec \nabla  \times \vec A(\vec r)$.
In this way, one  easily finds:

\begin{subequations}
 \begin{equation}
 <\psi|H_s|\psi>=  -\vec \mu \cdot \vec B
 \end{equation}
\text{with}
\begin{equation}
 \vec \mu = e ~G^{(d)} \vec \sigma
\end{equation}
\end{subequations}
and
\begin{equation}\label{muequal}
 G^{(d)}={\frac {N^2} {E+m}}.
\end{equation}
where we have assumed that the wave function $|\varphi>$ is standardly normalized:
$<\varphi|\varphi>=1$.


In the case of our model, the expression of eq. (\ref{muequal}) cannot be used in a straightforward way,
for the following reason.
The one particle wave function is expressed as an expansion in the relativistic HO basis, as shown in eq. (\ref{expansion}).
From that expression one cannot determine analytically the total normalization constant $N$;
on the other hand, a numerical calculation of that quantity would be affected by  numerical uncertainties.
In consequence, to calculate the magnetic dipole operator, we prefer to follow a different procedure.

\noindent
For the calculation, we have in mind the case of the $N(939)$.
In consequence, we consider a state
with $n'=0$, $l=m_l=0$, denoted by $|\psi_0>$.
Starting from eq. (\ref{hint_s}), using the expansion of the Dirac wave function given by eq.  (\ref{expansion}), and also
eq. (\ref{psidir_ho}), with standard handlings, one finds:

$$  <\psi_{0}|H_s|\psi_{0}>= \vec \sigma \cdot  {\frac e {4\pi}} \sum^{n_{max}}_{n_a,n_b=0} a^*_{n_b} {a}_{n_a} N_{{n_e}_b}N_{{n_e}_a}$$ 
 
\begin{equation}\label{H_s}
  \cdot \int d^3 r (\hat{ \vec r} \times \vec A(\vec r)) \left( {\frac {R'_{n_b,0} R_{n_a,0} } {E_{{n_e}_b}+m}  }+ {\frac {R_{n_b,0} R'_{n_a,0} } {E_{{n_e}_a} +m} } \right)
  \end{equation}
where all the indices not relevant for the calculation have been dropped; for brevity
we have also dropped the argument of the radial wave functions: $R_{n,0}= R_{n,0}(r)$; finally, the apex denotes the derivative with respect to the
radial coordinate $r$.
For a uniform magnetic field $\vec B$, we set:
\begin{equation}
 \vec A(\vec r)= {\frac 1 2} \vec B \times \vec r~.
\end{equation}
By using standard vectorial identities and replacing, under spherical integration, 
$(\vec \sigma \cdot  \hat{ \vec r} )(\vec B  \cdot \hat{ \vec r} ) \rightarrow {\frac 1 3} \vec \sigma \cdot \vec B$,
one finally obtains:

\begin{subequations}
\begin{equation}
<\psi_{0}|H_s|\psi_{0}>= -\vec \mu \cdot \vec B 
\end{equation}
with
\begin{equation}
 \vec \mu = e ~G^{(d)}_0 \vec \sigma
\end{equation}
\end{subequations}
and
$$ G^{(d)}_0 =
- {\frac 1 3}\sum^{n_{max}}_{n_a,n_b=0} a^*_{n_b} {a}_{n_a} N_{{n_e}_b}N_{{n_e}_a} $$
\begin{equation}\label{gd0}
\cdot \int_0^{\infty} dr ~r^3 \left( {\frac {R'_{n_b,0} R_{n_a,0} } {E_{{n_e}_b}+m}  }+ {\frac {R_{n_b,0} R'_{n_a,0} } {E_{{n_e}_a} +m} } \right) ~.
\end{equation}
We recall again that this expression has been derived for the case of $l=0$. Taking only one term in the expansion of the wave fuction, 
one  recovers, with standard handling, the expression 
of eq. (\ref{muequal}).
The single particle spatial matrix element of eq. (\ref{gd0}) is used to calculate the magnetic moment of the $N(939)$.

\section{ Wave functions and solutions of the Hamiltonian equation }\label{app_wf}
We now specify the form of the total wave functions of the model.
To this aim we take into account the coupling scheme discussed in sect. \ref{general} and synthetized in eq. (\ref{psi}).
We start with the two-component (Pauli) wave function, that (omitting the color factor) can be written by means 
of the  \textit{four factors} given in the following. 

\noindent
i) We start with the \textit{radial factor}:
\begin{equation}
 {\cal R}_{n_r,L}= R_{0,0}(r_1)R_{0,0}(r_2)R_{n_r,L}(r_3)~.
\end{equation}
where the first two terms correspond to the quarks 1 and 2; the third term corresponds to the quark 3; 
 for the ground states, one has  $n_r=0,~L=0$;

\noindent
ii) The \textit{angular factor} is:
\begin{equation}
{\cal Y}_{L,M_L}= Y_{0,0}(\hat r_1) Y_{0,0}(\hat r_2) Y_{L,M_L}(\hat r_3)= {\frac {1} {4\pi}} Y_{L,M_L}(\hat r_3)~;
\end{equation}
one has $Y_{0,0}(\hat r_3)={\frac {1} {\sqrt{4\pi}}}$ for the ground states.

\noindent
iii) The \textit{spin factor} has the form:
\begin{equation}\label{spin_factor}
 {\cal X}^{S_c}_{S,M_S}= [[\chi_{1/2}(1) \otimes  \chi_{1/2}(2)]_{S_c}  \otimes \chi_{1/2}(3)]_{S,M_S}~.
\end{equation}

\noindent
iv) Analogously, the \textit{isospin factor} is:
\begin{equation}\label{isospin_factor}
 {\cal P}^{T_c}_{T,M_T}  = [[\phi_{1/2}(1) \otimes  \phi_{1/2}(2)]_{T_c}  \otimes \phi_{1/2}(3)]_{T,M_T}~.
\end{equation}
In eqs. (\ref{spin_factor}) and (\ref{isospin_factor}) $S_c$ and $T_c$ respectively represent the spin and the isospin quantum numbers of the core.

\noindent
For the $N(939)$ (ground state, with $S=T=1/2$) the total wave function can be written in the form:
\begin{subequations}
\begin{equation} \label{Phi_N_a}
 \Phi_N= {\cal R}_{0,0} \cdot {\cal Y}_{0,0} \cdot 
 {\cal Q}^{(N)}_{M_S M_T}
\end{equation}
with
\begin{equation} \label{Phi_N_b}
 {\cal Q}^{(N)}_{M_S, M_T}= 
  {\frac {1} {\sqrt{2}}} 
  \left[   {\cal X}^{0}_{1/2,M_S} {\cal P}^{0}_{1/2,M_T}  +   {\cal X}^{1}_{1/2,M_S} {\cal P}^{1}_{1/2,M_T}     \right]~.
\end{equation}
\end{subequations}
Note that the spin-isospin factor of eq. (\ref{Phi_N_b})   has the same form as  the corresponding factor of 
the  CQMs and  is \textit{completely symmetric} with respect to quark interchange. 

\noindent
For the $\Delta(1232)$, one has:
\begin{equation}\label{psi_gen_fact}
 \Phi_{\Delta}=  {\cal R}_{0,0} \cdot {\cal Y}_{0,0} \cdot  {\cal X}^{1}_{3/2,M_S} \cdot {\cal P}^{1}_{3/2,M_T}
\end{equation}
that is also  \textit{completely symmetric}.

\noindent
For the excited states we have:
\begin{equation}
\Phi_{E}=  {\cal R}_{n_r,L} \cdot{\cal Y}_{L,M_L}\cdot {\cal X}^{S_c}_{S,M_S} \cdot {\cal P}^{T_c}_{T,M_T} ~.
\end{equation}
The Dirac wave function is constructed  applying to these functions the Dirac operators 
introduced in eq. (\ref{psidirb}):
\begin{equation}\label{Psi_lambda}
\Psi_{\Lambda}=D_1 D_2 D_3 \Phi_{\Lambda}~,
\end{equation}
where $\Lambda$ stands for $N, \Delta$ and $ E$; also $D_i=  D(\vec p_i\cdot\vec \sigma_i; E_i, m_q) $.

\vskip 0.5 truecm
\noindent
The one-body Dirac equation is solved analytically for the harmonic interaction of eq. (\ref{ugen}).
With these harmonic oscillator eigenfunctions, we calculate the matrix elements of the interaction $U^{(1)}(r_i)$.
We also add the spin-isospin dependent interaction of eqs. (\ref{hstia}) and  (\ref{hstib}).
Then, we diagonalize the total Hamiltonian matrix obtaining the approximate eigenvalues and eigenfunctions of the relativistic equation.
For each resonance, we take $10$ oscillator eigenfunctions.

Finally, we calculate perturbatively the contributions of $H_x$ with the nonrelativistic  expansion of eq.(\ref{k_x_nr}).
Due to the single particle character of the model, the total mass of each resonance is obtained summing the contributions of the three quarks.



\begin{thebibliography}{50} 
\bibitem{bag1}
A. Chodos, R. L. Jaffe, K. Johnson, C.
 B. Thorn, and V. F. Weisskopf,
Phys.\ Rev.\ D {\bf 9}, 3471 (1974).

\bibitem{bag2}
A. Chodos, R. L. Jaffe, K. Johnson,
 and C. B. Thorn,
Phys.\ Rev.\ D {\bf 10}, 2599 (1974).

\bibitem{bag3}
T. DeGrand, R. L. Jaffe, K. Johnson, and J. Kiskis,
Phys.\ Rev.\ D {\bf 12}, 2060 (1975).

\bibitem{bag4}
K. Johnson,
Acta Phys. Polonica {\bf B6}, 865 (1975).

\bibitem{bag5}
A. Bernotas and V. Simonis,
Nucl. Phys.{\bf A 741}, 179 (2004). 

\bibitem{bag6}
Sh. Mamedov,
 Eur. Phys.J. {\bf C30} 583 (2003).

\bibitem{bag7}
K. Colanero and M.-C. Chu,
Phys. Rev. C {\bf 65}  045203 (2002).

\bibitem{bag8}
K. Colanero and M.-C. Chu,
J. Phys. {\bf A 35}, 993 (2002). 

\bibitem{bag9}
D.Tadic and S. Zganec,
Phys. Rev. D {\bf 50}, 5853 (1994). 

\bibitem{bag10}
Xing Min Wang and Xiaotong Song,
 Phys.Rev. C {\bf 51}, 2750 (1995). 
\bibitem{chir1}
E.M. Tursunov,
\textit{ A periodic table for the excited Nucleon and Delta spectrum in a relativistic chiral quark model },
(hep-ph) arXiv:1103.3661 (2011).

\bibitem{chir2}
E.M. Tursunov  and  S. Krewald,
Phys. Rev. D {\bf 90}, 074015 (2014) 

\bibitem{isg:1}
  N.~Isgur and G.~Karl,
  Phys.\ Rev.\ D {\bf 18}, 4187 (1978).
\bibitem{isg:2}
  N.~Isgur and G.~Karl,
 Phys.\ Rev.\ D
  {\bf 19}, 2653 (1979).
\bibitem{isg:3}
N.~Isgur and G.~Karl,
Phys.\ Rev.\ D
  {\bf 20}, 1191 (1979).


\bibitem{Capstick:1986bm}
  S.~Capstick and N.~Isgur,
  Phys.\ Rev.\ D {\bf 34}, 2809 (1986).


\bibitem{bij:1}
  R.~Bijker, F.~Iachello and A.~Leviatan,
  Annals Phys.\  {\bf 236}, 69 (1994).
\bibitem{bij:2}
R.~Bijker, F.~Iachello and A.~Leviatan,
 Annals Phys.\    {\bf 284}, 89 (2000).


\bibitem{blr:1n1}
  M.~Ferraris, M.~M.~Giannini, M.~Pizzo, E.~Santopinto and L.~Tiator,
  Phys.\ Lett.\ B {\bf 364}, 231 (1995).
  
\bibitem{sig}
E. Santopinto, F. Iachello and M.M. Giannini,
Eur. Phys. J. A {\bf 1}, 307(1998).
\bibitem{bis}
R. Bijker, F. Iachello and E, Santopinto,
J. Phys. A {\bf 31}, 9041 (1998).
%
\bibitem{blr:1n2}
  M.~M.~Giannini, E.~Santopinto and A.~Vassallo,
  Eur.\ Phys.\ J.\ A {\bf 12}, 447 (2001).
\bibitem{blr:1n3}
  E.~Santopinto, A.~Vassallo, M.~M.~Giannini and M.~De Sanctis,
  Phys.\ Rev.\ C {\bf 76}, 062201 (2007).
\bibitem{blr:1n4}
E.~Santopinto, A.~Vassallo, M.~M.~Giannini and M.~De Sanctis,
  Phys.\ Rev.\ C {\bf 82}, 065204 (2010).

\bibitem{blr:2n1}
  L.~Y.~ Glozman and D.~O.~Riska,
  Phys.\ Rept.\  {\bf 268}, 263 (1996).
\bibitem{blr:2n2}
  L.~Y.~ Glozman, W.~Plessas, K.~Varga and R.~F.~Wagenbrunn,
  Phys.\ Rev.\ D {\bf 58}, 094030 (1998).
\bibitem{blr:2n3}
  R.~F.~Wagenbrunn, S.~Boffi, W.~Klink, W.~Plessas and M.~Radici,
  Phys.\ Lett.\  B {\bf 511}, 33 (2001).
\bibitem{blr:2n4}
  S.~Boffi, L.~Y.~Glozman, W.~Klink, W.~Plessas, M.~Radici and R.~F.~Wagenbrunn,
  Eur.\ Phys.\ J.\  A {\bf 14}, 17 (2002).


\bibitem{lor:1}
U.~Loring, K.~Kretzschmar, B.~C.~Metsch and H.~R.~Petry,
Eur.\ Phys.\ J.\ A {\bf 10}, 309 (2001).

\bibitem{lor:2}
U.~Loring, K.~Kretzschmar, B.~C.~Metsch and H.~R.~Petry,
 Eur.\ Phys.\ J.\ A {\bf 10}, 395 (2001).
  

\bibitem{Nakamura:2010zzi}
  K.~A.~Olive {\it et al.}  [Particle Data Group Collaboration],
  Chin.\ Phys.\ C {\bf 38}, 090001 (2014). 

\bibitem{Capstick:1992uc} 
  S.~Capstick,
  Phys.\ Rev.\ D {\bf 46}, 2864 (1992).
  
\bibitem{Capstick:1992th} 
  S.~Capstick and W.~Roberts, Phys. Rev. D 
  {\bf 58}, 074011 (1998).


\bibitem{blr:3n1}
  V.~Cred\'e {\it et al.}  [CB-ELSA Collaboration],
  Phys.\ Rev.\ Lett.\  {\bf 94}, 012004 (2005).
\bibitem{blr:3n2}
  D.~Trnka {\it et al.}  [CBELSA/TAPS Collaboration],
  Phys.\ Rev.\ Lett.\  {\bf 94}, 192303 (2005).

\bibitem{blr:4n1} 
  B.~Krusche {\it et al.},
  Phys.\ Rev.\ Lett.\  {\bf 74}, 3736 (1995).
\bibitem{blr:4n2}
  F.~Harter {\it et al.},
  Phys.\ Lett.\ B {\bf 401}, 229 (1997).
\bibitem{blr:4n3}
  M.~Wolf {\it et al.},
  Eur.\ Phys.\ J.\ A {\bf 9}, 5 (2000).


\bibitem{graal:1}
  F.~Renard {\it et al.}  [GRAAL Collaboration],
  Phys.\ Lett.\ B {\bf 528}, 215 (2002).
\bibitem{graal:2}
  Y.~Assafiri {\it et al.}  [GRAAL Collaboration], 
  Phys.\ Rev.\ Lett.\  {\bf 90}, 222001 (2003).





\bibitem{saphir:1}
  M.~Q.~Tran {\it et al.}  [SAPHIR Collaboration],
  Phys.\ Lett.\ B {\bf 445}, 20 (1998).
\bibitem{saphir:2}
  K.~H.~Glander {\it et al.} [SAPHIR Collaboration],
  Eur.\ Phys.\ J.\ A {\bf 19}, 251 (2004).



\bibitem{clas:1}
  M.~Dugger {\it et al.}  [CLAS Collaboration],
  Phys.\ Rev.\ Lett.\  {\bf 89}, 222002 (2002).
\bibitem{clas:2}
  M.~Dugger {\it et al.}  [CLAS Collaboration],
 Phys.\ Rev.\ Lett.\   {\bf 96}, 062001 (2006).
\bibitem{clas:3}
        M.~Ripani {\it et al.}  [CLAS Collaboration],
  Phys.\ Rev.\ Lett.\  {\bf 91}, 022002 (2003).



\bibitem{blr:5n1}
  M.~Ida and R.~Kobayashi,
  Prog.\ Theor.\ Phys.\  {\bf 36}, 846 (1966).
\bibitem{blr:5n2}
  D. B. Lichtenberg and L. J. Tassie,
  Phys. Rev. {\bf 155}, 1601 (1967).

\bibitem{Anselmino:1992vg}
  M.~Anselmino, E.~Predazzi, S.~Ekelin, S.~Fredriksson and D.~B.~Lichtenberg,
  Rev.\ Mod.\ Phys.\  {\bf 65}, 1199 (1993).
  
\bibitem{Jaffe:2004ph}  
  R.~L.~Jaffe,
  Phys.\ Rept.\ {\bf 409}, 1 (2005) hep-ph/0409065.


\bibitem{Wilczek:2004im}  
  F.~Wilczek, in \textit{From Fields to Strings},
Edited by M. Shifman, A. Vainshtein and J. Wheater.
Published by World Scientific Publishing Co. Pte. Ltd., 2005.
 ISBN: 978-981-4482-04-2 (ebook),
vol. 1, pp 77-93.  
  
\bibitem{Selem:2006nd}
  A.~Selem and F.~Wilczek,
in \textit{New trends in HERA physics. Proceedings, Ringberg Workshop, Tegernsee, Germany, October 2-7, 2005.}
Edited by G. Grindhammer, W. Ochs, B A Kniehl and G. Kramer. Published by World Scientific Publishing Co. Pte. Ltd., 2006. 
ISBN: 978-9812-773-52-4  
pp. 337-356 hep-ph/0602128.


\bibitem{Santopinto:2004hw}
  E.~Santopinto,
  Phys.\ Rev.\  C {\bf 72}, 022201 (2005).
  
\bibitem{Forkel:2008un}
  H.~Forkel and E.~Klempt,
  Phys.\ Lett.\  B {\bf 679}, 77 (2009).  


\bibitem{anis:1}
  A.~V.~Anisovich, V.~V.~Anisovich, M.~A.~Matveev, V.~A.~Nikonov, A.~V.~Sarantsev and
  T.~O.~Vulfs,
  Int.\ J.\ Mod.\ Phys.\  A {\bf 25}, 2965 (2010).
\bibitem{anis:2}
  A.~V.~Anisovich, V.~V.~Anisovich, M.~A.~Matveev, V.~A.~Nikonov, A.~V.~Sarantsev and
  T.~O.~Vulfs,
  Int.\ J.\ Mod.\ Phys.\  A {\bf 25}, 3155 (2010).

\bibitem{ferr:1}
  J.~Ferretti, A.~Vassallo and E.~Santopinto,
  Phys.\ Rev.\  C {\bf 83}, 065204 (2011).
\bibitem{ferr:2}
  E.~Santopinto and J.~Ferretti,
  Phys.\ Rev.\ C {\bf 92}, 025202 (2015).

\bibitem{DeSanctis:2011zz}
  M.~De~Sanctis, J.~Ferretti, E.~Santopinto and A.~Vassallo,
  Phys.\ Rev.\  C {\bf 84}, 055201 (2011).  
  
\bibitem{Galata:2012xt} 
  G.~Galat\`a and E.~Santopinto,
  Phys.\ Rev.\ C {\bf 86}, 045202 (2012). 
  
\bibitem{col_last}
M. De Sanctis, J. Ferretti, E. Santopinto and A. Vassallo,
Eur. Phys. J. A  {\bf 52}, 121 (2016).

  
  
 
 \bibitem{GellMann:1964nj}
  M.~Gell-Mann,
  Phys.\ Lett.\  {\bf 8}, 214 (1964).
  
\bibitem{Jakob:1997}
        R.~Jakob, P.~J.~Mulders and J.~Rodrigues,
  Nucl.\ Phys.\  A {\bf 626}, 937 (1997).
  
\bibitem{Bloch:1999ke}
 J.~C.~R.~Bloch, C.~D.~Roberts, S.~M.~Schmidt, A.~Bender and M.~R.~Frank,
 Phys.\ Rev.\  C {\bf 60}, 062201 (1999).  
  
\bibitem{Brodsky:2002}
        S.~J.~Brodsky, D.~S.~Hwang and I.~Schmidt,
  Phys.\ Lett.\  B {\bf 530}, 99 (2002).  
  
\bibitem{Ma}
        B.~Q.~Ma, D.~Qing and I. Schmidt, 
        Phys. Rev. C {\bf 65}, 035205 (2002).    
    
\bibitem{Oettel:2002wf}
  M.~Oettel and R.~Alkofer,
  Eur.\ Phys.\ J.\  A {\bf 16}, 95 (2003).  
  
\bibitem{Gamberg:2003}
  L.~P.~Gamberg, G.~R.~Goldstein and K.~A.~Oganessyan,
  Phys.\ Rev.\  D {\bf 67}, 071504 (2003). 
  
\bibitem{Jaffe:2003sg} 
  R.~L.~Jaffe and F.~Wilczek,
  Phys.\ Rev.\ Lett.\  {\bf 91}, 232003 (2003).  
  
\bibitem{Maris:2004}
  P.~Maris,
  Few Body Syst.\  {\bf 35}, 117 (2004).  
   
\bibitem{DeGrand:2007vu}
  T.~DeGrand, Z.~Liu and S.~Schaefer,
  Phys.\ Rev.\  D {\bf 77}, 034505 (2008).
  
\bibitem{BacchettaRadici}
        A.~Bacchetta, F.~Conti and M.~Radici,
  Phys.\ Rev.\  D {\bf 78}, 074010 (2008).  

\bibitem{pf1}
 W.~H.~Klink,
  Phys.\ Rev.\  C {\bf 58}, 3587 (1998).
\bibitem{pf2}
 W.~H.~Klink, 
  Phys.\ Rev.\  C {\bf 58}, 3617 (1998).
\bibitem{pf3}
   W.~N.~Polyzou {\it et al.},
  Few Body Syst.\  {\bf 49}, 129 (2011).
 
  
    



  



 

 

\bibitem{pom1}
 O. Nachtmann, 
 Contribution to the Ringberg Workshop on HERA Physics 2003,
 Report Number:  HD-THEP-03-63;
 arXiv:hep-ph/0312279

\bibitem{pom2}
Lon-chang Liu,
Nucl. Phys. {\bf A755}, 607 (2005).

\bibitem{pom3}
C. Ewerz, P. Lebiedowicz, O. Nachtmann and A. Szczurek,
Physics Letters B {\bf 763}, 382 (2016). 
 
\bibitem{pom4}
M. N. Sergeenko,
Gluonium states and the Pomeron trajectory,
arXiv:0807.0911v1 [hep-ph] (2008).

\bibitem{pdg} 
 M. Tanabashi \textit{et al.},
(Particle Data Group )
 Phys. Rev. D {\bf 98}, 030001 (2018).






\bibitem{spsya}
 J. S. Bell and H. Ruegg, Nucl. Phys.
{\bf B 98}, 151 (1975).

\bibitem{spsyb}
 J. S. Bell and H. Ruegg, (Errata),  Nucl. Phys.
{\bf B104}, 546 (1976).
 
\bibitem{spsyc}
G. B. Smith and L. J. Tassie, Annals of Phys. {\bf 65}, 352 (1971).

\bibitem{spsyd}
Richard L. Hall and Petr Zorin,
Eur. Phys. J. Plus {\bf 131}, 102 (2016).

\bibitem{spsye}
J. N. Ginocchio, Phys. Rep. {\bf 414}, 165 (2005).
 
\bibitem{spsyf} 
P. Alberto , A. Castro, M. Fiolhais, R. Lisboa and M. Malheiro,
Journal of Phys. (Conference Series), {\bf 490} 012069 (2014).
 
\bibitem{spsyg}
P. Alberto, A. S. de Castro, and M. Malheiro,
Phys. Rev. {\bf C 87}, 031301(R) (2013).


\bibitem{pdg_old}
 C. Patrignani  \textit{et al.},
(Particle Data Group )
 Chin. Phys. C, {\bf 40}, 100001 (2016) and 2017 update.


 
\bibitem{mdes_edep}
M. De Sanctis, Cent. Eur. J. Phys. {\bf12}, 221 (2014)



\end{thebibliography}
\end{document}